\documentclass[onecolumn]{aa}		

\usepackage{graphicx,txfonts,natbib}
\bibpunct{(}{)}{;}{a}{}{,}              

\newcommand{\msun}{\hbox{M$_{\odot}$}}

\newcommand{\otwo}{\hbox{[O II]$\lambda 3727$}}
\newcommand{\nethree}{\hbox{[Ne III]$\lambda 3869$}}

\newcommand{\othreea}{\hbox{[O III]$\lambda 4363$}}
\newcommand{\hbeta}{\hbox{H$\beta$}}
\newcommand{\othree}{\hbox{[O III]$\lambda\lambda 4959,5007$}}

\newcommand{\othreec}{\hbox{[O III]$\lambda 5007$}}

\newcommand{\halpha}{\hbox{H$\alpha$}}
\newcommand{\ntwob}{\hbox{[N II]$\lambda 6583$}}
\newcommand{\ntwo}{\hbox{[N II]$\lambda\lambda 6548,6583$}}
\newcommand{\ntwootwo}{\hbox{[N II]/[O II]}}
\newcommand{\stwo}{\hbox{[S II]$\lambda\lambda 6716,6731$}}
\newcommand{\stwoa}{\hbox{[S II]$\lambda 6716$}}
\newcommand{\stwob}{\hbox{[S II]$\lambda 6731$}}

\begin{document}

\title{
Nebular Abundances of \\ Nearby Southern Dwarf Galaxies 
}

\author{
Henry Lee$\,$\inst{1,2}\fnmsep\thanks{
Visiting Astronomer, Cerro Tololo Inter-American Observatory,
National Optical Astronomy Observatories, operated by the
Association of Universities for Research in Astronomy, Inc. (AURA)
under cooperative agreement with the National Science Foundation.
},
Eva K. Grebel$\,$\inst{1}, \and
Paul W. Hodge$\,$\inst{3}
}

\institute{
Max-Planck-Institut f\"ur Astronomie, K\"onigstuhl 17, D-69117
Heidelberg, Germany.
\and
Dept. of Physics \& Astronomy, York University,
4700 Keele St., Toronto, Ontario  M3J~1P3  Canada.
\and
Astronomy Department, University of Washington,
Box 351580, Seattle, WA  98195-1580 USA.
}

\abstract{				
The results of optical spectroscopy of \ion{H}{II} regions
in a sample of southern dwarf irregulars consisting of
five dwarf galaxies in the Centaurus~A group, four dwarfs in the
Sculptor group, and eight additional dwarf galaxies are presented. 
Oxygen abundances are derived using the direct method where \othreea\
is detected; otherwise, abundances are derived with the bright-line
method using the McGaugh and the Pilyugin calibrations.
ESO358$-$G060 has the lowest oxygen abundance (12$+$log(O/H) = 7.32),
which is comparable to the value for the second most metal-poor galaxy
known (SBS~0335$-$052).
In all, new oxygen abundances are reported for nine dwarf galaxies;
updated values are presented for the remaining galaxies.
Derived oxygen abundances are in the range from 3\% to 26\%
of the solar value.
Oxygen abundances for dwarfs in the southern sample are consistent 
with the metallicity-luminosity relationship defined by
a control sample of dwarf irregulars with \othreea\ abundances and
well-measured distances.
However, NGC~5264 appears to have an (upper branch) oxygen abundance
approximately two to three times higher than other dwarfs at similar
luminosities. 
Nitrogen-to-oxygen and neon-to-oxygen abundance ratios are
also reported; in particular, IC~1613 and IC~5152 show elevated
nitrogen-to-oxygen ratios for their oxygen abundances.
}

\date{Received 30 August 2002 / Accepted 15 January 2003}
\offprints{Henry Lee; \email{lee@mpia.de}}
\titlerunning{Nebular Abundances of Southern Dwarfs}
\authorrunning{Lee et al.}

\maketitle

\section{Introduction}

Dwarf galaxies, the most abundant type of galaxy in groups and 
clusters, may hold the key to understanding galaxy evolution.  
In standard hierarchical structure formation scenarios, they are the 
building blocks of more massive galaxies.  
Conversely, the environment -- e.g., the presence or absence of more
massive galaxies in the vicinity -- surrounding dwarf galaxies is
expected to affect their evolution (e.g., \citealp{mayer01}).  
Furthermore, the intrinsic properties of a dwarf galaxy -- e.g., its
mass, density, gas content -- are likely to influence the
histories of star formation and chemical enrichment.
However, the relative importance of external and internal properties
and their actual effect on dwarf galaxy evolution remains unknown.   
The most detailed studies of dwarf galaxy properties have been carried
out in the Local Group, but the limited sample size and the wide range
of properties prevent us from obtaining a unified picture.  

To arrive at a better understanding of what drives dwarf galaxy 
evolution we are carrying out a comprehensive multi-wavelength 
study of dwarf galaxies in the Local Volume ($\la 5$~Mpc).
The Local Volume contains several nearby galaxy groups as well as 
galaxies in the relative isolation of the field.  
After the conclusion of an all-sky survey to detect dwarf galaxy
candidates in the Local Volume 
\cite[][and references therein]{kara00}
we are carrying out ground-based imaging to derive integrated
photometry and structural parameters (e.g., \citealp{makarova02}).
These efforts are complemented by two snapshot programs with the Hubble
Space Telescope's (HST) Wide Field and Planetary Camera 2 (WFPC2)
totalling 200 orbits to study stellar content, recent star formation
histories, distances, and relative positions within galaxy groups
(e.g., \citealp{kara02a,kara02b,kara02c}).
Radio observations (e.g., \citealp{hucht00a,hucht00b}) reveal the
\ion{H}{I} content and radial velocity of the dwarfs, provided
that they contain detectable amounts of \ion{H}{I}.  
For a description of the project as a whole, see \cite{grebel00}.

Since our HST data enable us to measure relative distances within
galaxy groups, we are in the unique position of being able to 
assign membership probabilities to individual dwarf galaxies and
to consider them based on their three-dimensional location within
a group.  
Our ground-based imaging in combination with distances yields
information on the luminous mass of dwarf galaxies.   
Compiling a database that, in addition, contains accurate information
on chemical abundances will allow us to consider dwarf galaxies within
their evolutionary context in groups and in the field.  
The knowledge of relative distances and intrinsic properties is a
major step towards resolving what governs dwarf galaxy evolution.

The data presented here are part of our continuing efforts to obtain
abundances for dwarf irregular galaxies (dIs) in the Local Volume and
are based on the first two observing runs dedicated to this multi-year
project. 
However, not all targets have HST-based distances.
At the time of writing, about 150 out of the 200 dwarf galaxy
candidates in the Local Volume have been observed, as we await the
completion of the HST snapshot survey.
Galaxies were selected according to the following criteria:
(a) galaxies were visible at the epoch of observations, and
(b) galaxies were sufficiently luminous to be observable with
a 1.5-m class telescope.
Hence, the present study consists primarily of results for individual
galaxies and of the discussion for these galaxies in the
metallicity-luminosity relationship.
The solar value of the oxygen abundance of 12$+$log(O/H) = 8.87
\citep{zsolar} is adopted for the present work to facilitate
comparisons with earlier studies.
However, recent work has indicated that the solar value may in fact 
be smaller by about 0.1 to 0.2~dex \citep{zsolarnew,holweger01}.

The outline of this paper is as follows.
A brief description of the observed galaxies is presented in
Sect.~\ref{sec_sample}. 
Observations and reductions of the data are presented in
Sect.~\ref{sec_obs}.
Measurements and analysis are discussed in Sect.~\ref{sec_analysis},
and nebular abundances are presented in Sect.~\ref{sec_abundderive}.
A discussion of individual galaxies and of the metallicity-luminosity
relation is provided in Sect.~\ref{sec_discuss}. 
A summary is given in Sect.~\ref{sec_concl}.

\section{The Samples of Dwarf Galaxies}
\label{sec_sample}

\subsection{The Southern Sample of dIs}

The southern sample consists of five dwarf galaxies from the 
Centaurus~A (NGC~5128, Cen~A) group, four dwarf galaxies from the 
Sculptor (Scl) group, one dwarf in the Antlia-Sextans group, and seven
dwarf galaxies in the Local Group and in the field. 
In particular, IC~1613, IC~5152, NGC~2915, and NGC~3109 were chosen
because oxygen abundances in the literature were derived from
measurements using obsolete technology or abundances were not
well measured.
Two galaxies (A0355$-$465, IC~2032) in the field were
selected from the list compiled by \cite{kk98}, and
two galaxies (ESO358$-$G060, ESO302$-$G014) were selected
from the \cite{fg85} catalog.
ESO358$-$G060 is a member of the Fornax Cluster; see
Sect.~\ref{sec_eso358g060} for additional details.

Because half of the galaxies are members of two nearby groups
of galaxies, a brief summary of prior spectroscopy 
is mentioned here.
The Centaurus~A group and the Sculptor group are each at a distance of
$\ga 5$~Mpc.
\cite{ws83} and \cite{webster83} obtained oxygen abundances for
southern irregular and spiral galaxies, including galaxies in the Cen~A
group and in the field, with a measured range of abundances from about
ten to sixty per cent of the solar value.
\cite{miller96} carried out \halpha\ imaging of eight Sculptor group
dwarfs; only two contained detectable \ion{H}{II} regions.
Subsequent \ion{H}{II} region spectroscopy of these two dwarfs showed
that their oxygen abundances were roughly 0.1 of the solar value,
although the temperature-sensitive \othreea\ emission line 
(see Sect.~\ref{sec_abundderive}) was not detected in either galaxy.
\cite{hgmo01} measured \othreea\ in the low surface brightness 
dI ESO245$-$G005 in the Scl group. 
\othreea\ was detected in two \ion{H}{II} regions; however, 
the two resulting oxygen abundances differ by 0.21~dex, which may
be evidence of an abundance gradient.
This is similar to the gradient ($\approx 0.2$ dex~kpc$^{-1}$) from
spectra of seven \ion{H}{II} regions for the same galaxy measured by
\cite{miller96}, although he determined oxygen abundances
using the indirect or bright-line method 
(see Sect.~\ref{sec_brightline}).

Properties of galaxies in the southern sample are listed in
Table~\ref{table_gxylist}.
Galaxies with probable Local Group membership are listed in
\cite{grebel99,grebel00rev}, and \cite{grebel03}.
Compilations of oxygen abundances for nearby dwarf irregulars
are found in, e.g., \cite{mateo98}, \cite{pilyugin01b}, and
\cite{lee03}.
Comments about individual galaxies from the southern sample will be
addressed in Sect.~\ref{sec_individ}.
\begin{table}
\scriptsize 
\begin{center}
\renewcommand{\arraystretch}{1.}
\caption{
Basic properties of galaxies in the southern sample.
Galaxies are grouped by association to a group (or otherwise)
and are listed in alphabetical order within each category.
All properties are obtained from NED, unless otherwise noted.
Cols.~(1) and (2): Galaxy name used in the present work; other
names. 
Col.~(3): Membership.
Col.~(4): Classification type.
Col.~(5): Heliocentric velocity.
Col.~(6): Total apparent $B$ magnitude.
Col.~(7): Total \ion{H}{I} 21-cm flux.
Rotational velocities in \ion{H}{I} have been measured for
A1243$-$335, A1334$-$277, DDO~161, IC~1613, and UGCA~442 
\citep{hoffman96,cote00}; a rotation curve in \halpha\
for ESO358$-$G060 has been measured by \cite{mg02}.
Col.~(8): Total \ion{H}{I} mass to blue luminosity ratio. 
Col.~(9): Measured or estimated distances.
Centaurus~A group galaxies: \cite{kara02b}, 
except DDO~161 \citep{cote00}.
Sculptor group galaxies: \cite{cote00}, 
except UGCA~442 (Grebel et al. 2003, in preparation).
Other southern galaxies: 
ESO358$-$G060 -- \cite{mould00}; 
IC~1613 -- \cite{dolphin01b};
IC~2032 -- \cite{carrasco01};
NGC~2915 -- \cite{kara03}; 
NGC~3109, Sag~DIG, and IC~5152 -- \cite{kara02c}.
NOTES: 
$^a$~From \cite{cote97}.
$^b$~From \cite{longmore82}.
$^c$~Velocity estimated from spectrum (Sect.~\ref{sec_individ}).
$^d$~From \cite{barnes97}.
$^e$~Member of Fornax Cluster Catalog \citep[FCC;][]{fcc}.
$^f$~From \cite{schroeder01}; see also \cite{matthews98}.
$^g$~From \cite{hoffman96}.
$^h$~From \cite{hucht00a}.
$^i$~From \cite{hr86}.
$^j$~From \cite{meurer96}.
$^k$~Common group of dwarfs including NGC~3109, Sextans~A,
Sextans~B, and the Antlia dwarf \citep{vdb99,tully02}.
$^l$~From \cite{bdb01}.
$^m$~From \cite{lk00}.
$^n$~From \cite{yl97}.
%
\label{table_gxylist}
}
\begin{tabular}{lcccccccc}
\hline \hline
Galaxy & Other & Member & Type & $v_{\odot}$ & $B_T$ & $F_{21}$ & 
$M_{\rm HI}/L_B$ & $D$ \\
& Name(s) & & & (km s$^{-1}$) & (mag) & (Jy km s$^{-1}$) & 
$(M_{\odot}/L_{{\odot},B})$ & (Mpc) \\
(1) & (2) & (3) & (4) & (5) & (6) & (7) & (8) & (9) \\
\hline
\multicolumn{8}{c}{{\sf Centaurus A group dwarfs}} \\
\hline
\object{A1243$-$335} & \object{ESO381$-$G020}, \object{AM1243$-$333} 
& Cen A & 
IB(s)m & $+585$ & 14.24 & 31.9$\,^a$ & 1.85 & 3.63 \\
\object{A1324$-$412} & \object{ESO324$-$G024}, \object{AM1324$-$411}
& Cen A & 
IAB(s)m: & $+513$ & 12.91 & 52.1$\,^a$ & 0.75 & 3.73 \\
\object{A1334$-$277} & \object{ESO444$-$G084}, \object{AM1334$-$274}
& Cen A & 
Im & $+587$ & 15.01 & 19.6$\,^a$ & 2.29 & 4.61 \\
\object{A1346$-$358} & \object{ESO383$-$G087}, \object{AM1346$-$354} 
& Cen A & 
SB(s)dm & $+326$ & 11.00 & 27.4$\,^a$ & 0.08 & 3.63 \\
\object{DDO 161} & \object{UGCA 320}, \object{A1300$-$17} & Cen A &
IB(s)m sp & $+744$ & 13.52 & 110.1$\,^a$ & 3.10 & 5.25 \\
%
\object{NGC 5264} & \object{DDO 242}, \object{UGCA 370} & Cen A &
IB(s)m & $+478$ & 12.60 & 13.7$\,^a$ & 0.19 & 4.53 \\
\hline
\multicolumn{8}{c}{{\sf Sculptor group dwarfs}} \\
\hline
\object{AM0106$-$382} & \ldots & Scl &
dwarf Im & $+645$ & 16.26 & $< 2.8\,^a$ & $<$ 1.29 & 3.0 \\
\object{ESO347$-$G017} & \object{PGC071466} & Scl &
SB(s)m: & $+659$ & 14.19 & 10.5$\,^a$ & 0.71 & 3.0 \\
\object{ESO348$-$G009} & \object{A2346$-$380}, \object{AM2346$-$380}
& Scl &
IBm & $+$657 & 13.60$\,^b$ & 8.4$\,^a$ & 0.33 & 3.0 \\
\object{UGCA 442} & \object{ESO471$-$G006}, \object{AM2341$-$321}
& Scl &
SB(s)m: & $+267$ & 13.60 & 54.3$\,^a$ & 2.12 & 4.27 \\
\hline
\multicolumn{8}{c}{{\sf Other southern dwarfs}} \\
\hline
\object{A0355$-$465} & \object{ESO249$-$G032}, \object{AM0355$-$463} 
& field & 
IB(s)m pec & $+1168\,^c$ & 16.35 & $< 0.20\,^b$ & $<$ 0.10 & 15.6 \\
\object{ESO302$-$G014} & \object{AM0349$-$383} & field & 
Im pec & $+881$ & 14.84 & 4.41$\,^d$ & 0.56 & \ldots \\
\object{ESO358$-$G060} & 
\object{FCC 302}$\,^e$ & Fornax Cluster & 
IB(s)m: & $+803$ & 15.86 & 12.27$\,^f$ & 3.97 & 20.0 \\
\object{IC 1613} & \object{DDO 8}, \object{UGC 668} & Local Group &
IB(s)m & $-$234 & 9.88 & 698$\,^g$ & 0.89 & 0.73 \\
\object{IC 2032} & \object{ESO156$-$G042}, \object{AM0405$-$552}
& Dorado & 
IAB(s)m pec: & $+1066\,^h$ & 14.78 & 2.87$\,^h$ & 0.35 & 17.2 \\
\object{IC 5152} & \object{ESO237$-$G027}, \object{AM2159$-$513}
& field &
IA(s)m & $+$124 & 11.03 & 98.0$\,^i$ & 0.35 & 2.07 \\
\object{NGC 2915} & \object{ESO037$-$G003}, \object{AM0926$-$762} & field & 
I0; BCD? & $+468$ & 12.93 & 145$\,^j$ & 1.10 & 3.78 \\
\object{NGC 3109} & \object{DDO 236}, \object{UGCA 194} 
& Antlia-Sextans$\,^k$ & 
SB(s)m & $+404$ & 10.26 & 1110$\,^l$ & 1.64 & 1.33 \\
%
\object{Sag DIG} & \object{ESO594$-$G004} & Local Group &
IB(s)m & $-$77 & 13.99$\,^m$ & 32.6$\,^n$ & 1.21 & 1.11 \\
\hline
\end{tabular}
\end{center}
\end{table}

\subsection{The Control Sample of dIs}
\label{sec_control}

\cite{rm95} constructed a sample of nearby dIs 
with direct (\othreea) oxygen abundances and well-measured distances
from resolved stellar photometry.
As a part of his thesis, Lee updated the sample incorporating 
updates from recent literature and unpublished spectroscopic 
data \citep{lee01,lee03}.
This sample of dIs will be referred to as the control sample, against
which the present sample of southern dwarfs will be compared.
Because new measurements of IC~1613 and NGC~3109 are discussed here,
they are excluded from the control sample in this paper and are
included in the present southern sample of dIs.

As part of a growing sample of galaxies which satisfy the criteria
above, UGC~4483 is added here to the list of dIs in
the control sample.
Direct oxygen and nitrogen abundances were measured by
\cite{skillman94} and \cite{itl94}.
Average values of 12$+$log(O/H) = 7.53 and log(N/O) = $-1.60$
are adopted here.
Tip of the red giant branch (TRGB) distances were obtained
independently by \cite{dolphin01a} and \cite{it02}; an average
distance of 3.3~Mpc ($m-M = 27.59$) is adopted here.
The integrated 21-cm flux integral is 13.34 Jy~km~s$^{-1}$
\citep{vanzee98}, which gives 
$M_{\rm HI} = 3.42 \times 10^7\,\msun$ and
$M_{\rm HI}/L_B = 1.96 \, M_{\odot}/L_{{\odot},B}$.

\section{Observations and Reductions}
\label{sec_obs}		

Long-slit spectroscopic observations were carried out in two
observing runs (2001 Feb. 2--4 UT and Aug. 17--20 UT) with the
Cassegrain Spectrograph on the 1.5-metre telescope at the Cerro
Tololo Inter-American Observatory. 
Details of the instrumentation employed and the log of observations
are listed in Tables~\ref{table_obsprops} and \ref{table_obslog},
respectively.
Observing conditions were varied.
For the February run, all three nights were clear, although
high thin patchy clouds were present at times.
The lunar illumination increased from 0.55 to 0.76.
For the August run, cloud and snow prevented observations
for the first two nights; the final two nights were clear
with near-zero lunar illumination.
The slit angle was kept constant in an east-west orientation
for all observations.
The central positions of the long slit for each target are listed 
in Table~\ref{table_obslog}.
After visual inspection of an image for each target,
the centre of the long-slit was placed at the 
brightest candidate \ion{H}{II} region for each galaxy.
\begin{table}
\scriptsize 
\begin{center}
\renewcommand{\arraystretch}{1.}
\caption{
Properties of Cassegrain Spectrograph employed at CTIO 1.5-m telescope.
NOTE: $^a$~Slit widths set for February and August 2001 runs,
respectively. 
%
\label{table_obsprops}
}
\begin{tabular}{lc}
\hline \hline
Property & Value \\ \hline
\multicolumn{2}{c}{{\sf Loral CCD}} \\ \hline
Total area & 1200 pix $\times$ 800 pix \\
Usable area & 1200 pix $\times$ 270 pix \\
Pixel size & 15 $\mu$m \\
Image scale & 1.3 arcsec pixel$^{-1}$ \\
Gain & 1.42 $e^-$ ADU$^{-1}$ \\ 
Read-noise (rms) & 6.5 $e^-$ \\ \hline
\multicolumn{2}{c}{{\sf Grating \#09}} \\ \hline
Groove density & 300 lines mm$^{-1}$ \\
Blaze $\lambda$ (1st order) & 3560 \AA \\
Dispersion & 2.87 \AA\ pixel$^{-1}$ \\
Effective $\lambda$ range & 3500--6950 \AA \\ \hline
\multicolumn{2}{c}{{\sf Long slit}} \\ \hline
Length & 5.8 arcmin \\
Width$\,^a$ & 2, 3 arcsec \\
\hline
\end{tabular}
\end{center}
\end{table}
\begin{table}
\scriptsize 
\begin{center}
\renewcommand{\arraystretch}{1.}
\caption{
Log of Observations.
Col.~(1): Galaxy name in alphabetical order (within each 
association).
Col.~(2): Date of observation.
Cols.~(3) and (4): Central position of the long-slit: 
right ascension in hours, minutes and seconds, and declination 
in degrees, minutes, and seconds (Epoch J2000). 
Col.~(5): Number of exposures obtained and the length of each
exposure in seconds. 
Col.~(6): Total exposure time.
Col.~(7): Mean effective airmass.
Col.~(8): \othreea\ : detected, upper limit, or not detected.
Col.~(9): Relative root-mean-square error in the sensitivity
function obtained from observations of standard stars.
NOTES:
$^a$~\ion{H}{II} region label from \cite{webster83}.
$^b$~\ion{H}{II} region number from \cite{miller96}.
$^c$~Only \othreec\ and \halpha\ detected.
$^d$~\ion{H}{II} region number from \cite{hlg90}; also known
as S3.
$^e$~\ion{H}{II} region label from \cite{talent80} and \cite{webster83}.
$^f$~\ion{H}{II} region number from \cite{rm92}.
$^g$~\ion{H}{II} region number from \cite{stm89} and \cite{strobel91}.
%
\label{table_obslog}
}
\begin{tabular}{lcccccccc}
\hline \hline
Galaxy & 2001 Date & 
$\alpha_{\rm slit}$ (J2000) & $\delta_{\rm slit}$ (J2000) & 
$N_{\rm exp}$ & $t_{\rm total}$ & $\langle X \rangle$ & 
\othreea & RMS \\ 
& (UT) & $(^h\;^m\;^s)$ & $(\degr\;\arcmin\;\arcsec)$ & & (s) & & & (mag) \\
(1) & (2) & (3) & (4) & (5) & (6) & (7) & (8) & (9) \\
\hline
\multicolumn{9}{c}{{\sf Centaurus A group dwarfs}} \\
\hline
A1243$-$335 (\ion{H}{II}\#A)$\,^a$ & 2 Feb & 
        12$\;$46$\;$02.4 & $-33\;50\;10.0$ &
        $12 \times 600$ & 7200 & 1.18 & \ldots & 3.5\% \\
A1324$-$412 & 4 Feb & 
        13$\;$27$\;$37.2 & $-41\;28\;29.4$ & 
        $2 \times 1200$ & 2400 & 1.47 & \ldots & 6.5\% \\
A1334$-$277 & 4 Feb & 
        13$\;$37$\;$20.3 & $-28\;02\;15.8$ & 
        $8 \times 900$ & 7200 & 1.13 & \ldots & 6.5\% \\
A1346$-$358 (\ion{H}{II}\#A)$\,^a$ & 3 Feb & 
        13$\;$49$\;$18.7 & $-36\;03\;10.4$ &
        $8 \times 900$ & 7200 & 1.08 & detected & 4.1\% \\
DDO 161 & 3 Feb & 
        13$\;$03$\;$16.7 & $-17\;25\;03.2$ & 
        $7 \times 900$ & 6300 & 1.36 & upper limit & 4.1\% \\
NGC 5264 & 2 Feb & 
        13$\;$41$\;$37.0 & $-29\;54\;29.0$ &
        $9 \times 600$ & 5400 & 1.05 & \ldots & 3.5\% \\
\hline
\multicolumn{9}{c}{{\sf Sculptor group dwarfs}} \\
\hline
AM0106$-$382 & 20 Aug & 
        01$\;$08$\;$24.9 & $-38\;11\;59.8$ & 
        $1 \times 300, 2 \times 1800$ & 3900 & 1.18 & \ldots & 5.0\% \\
ESO 347$-$G017 & 19 Aug & 
        23$\;$26$\;$56.3 & $-37\;20\;53.7$ & 
        $3 \times 1800$ & 5400 & 1.21 & upper limit & 5.0\% \\
ESO 348$-$G009 & 20 Aug & 
        23$\;$49$\;$26.5 & $-37\;45\;42.1$ & 
        $1 \times 300, 2 \times 1800$ & 3900 & 1.32 & \ldots & 5.0\% \\
UGCA 442 (\ion{H}{II}\#2)$\,^b$ & 19 Aug & 
        23$\;$43$\;$45.4 & $-31\;57\;27.4$ & 
	$1 \times 600, 4 \times 1800$ & 7800 & 1.07 & upper limit & 5.0\% \\
\hline
\multicolumn{9}{c}{{\sf Other southern dwarfs}} \\
\hline
A0355$-$465 (\ion{H}{II}\#B)$\,^a$ & 3 Feb & 
        03$\;$57$\;$22.5 & $-46\;22\;00.2$ &
        $6 \times 900$ & 5400 & 1.11 & upper limit & 4.1\% \\
ESO 302$-$G014$\,^c$ & 2 Feb & 
        03$\;$51$\;$40.9 & $-38\;27\;09.5$ &
        $1 \times 206, 4 \times 3600$ & 3806 & 1.17 & \ldots & 3.5\% \\
ESO 358$-$G060 & 4 Feb & 
        03$\;$45$\;$11.7 & $-35\;34\;22.9$ & 
        $7 \times 900$ & 6300 & 1.18 & upper limit & 6.5\% \\
IC 1613 (\ion{H}{II}\#13)$\,^d$ & 20 Aug & 
        01$\;$04$\;$47.3 & $+02\;06\;41.9$ & 
	$1 \times 300, 3 \times 1800$ & 5700 & 1.19 & upper limit & 5.0\% \\
IC 1613 (\ion{H}{II}\#37)$\,^d$ & 19 Aug & 
        01$\;$05$\;$00.8 & $+02\;04\;26.1$ & 
	$1 \times 600, 2 \times 1800$ & 4200 & 1.20 & detected & 5.0\% \\
IC 2032 & 2 Feb & 
        04$\;$07$\;$02.9 & $-55\;19\;31.2$ &
        $12 \times 600$ & 7200 & 1.46 & \ldots & 3.5\% \\
IC 5152 (\ion{H}{II}\#A)$\,^e$ & 20 Aug & 
        22$\;$02$\;$51.8 & $-51\;17\;12.0$ &
	$1 \times 300, 3 \times 1800$ & 5700 & 1.43 & detected & 5.0\% \\
NGC 2915 & 3 Feb & 
        09$\;$26$\;$28.7 & $-76\;37\;31.2$ & 
        $6 \times 900$ & 5400 & 1.47 & \ldots & 4.1\% \\
NGC 3109 (\ion{H}{II}\#6)$\,^f$ & 4 Feb & 
        10$\;$03$\;$04.2 & $-26\;09\;18.5$ &
        $6 \times 900$ & 5400 & 1.07 & \ldots & 6.5\% \\
Sag DIG (\ion{H}{II}\#3)$\,^g$ & 19 Aug & 
        19$\;$30$\;$02.3 & $-17\;41\;35.1$ &
	$6 \times 1800$ & 10800 & 1.05 & \ldots & 5.0\% \\
\hline
\end{tabular}
\end{center}
\end{table}

\begin{figure*} 
\centering
\includegraphics[width=\textwidth]{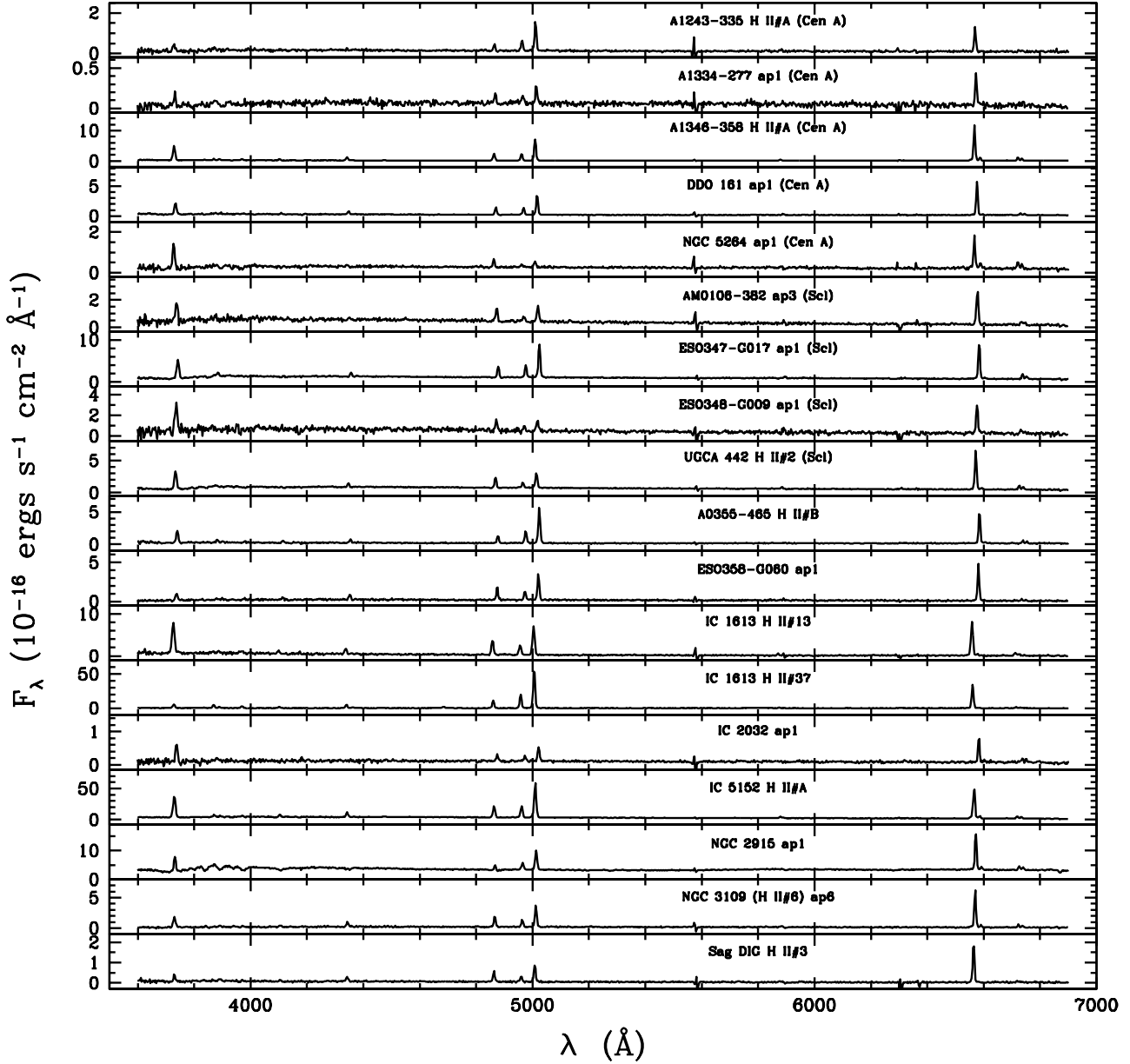} 
\caption{
Long-slit spectra.
The flux per unit wavelength is plotted against wavelength
in each panel.
}
\label{fig_allspc}
\end{figure*}

\begin{figure*} 
\centering
\includegraphics[angle=-90,width=12cm]{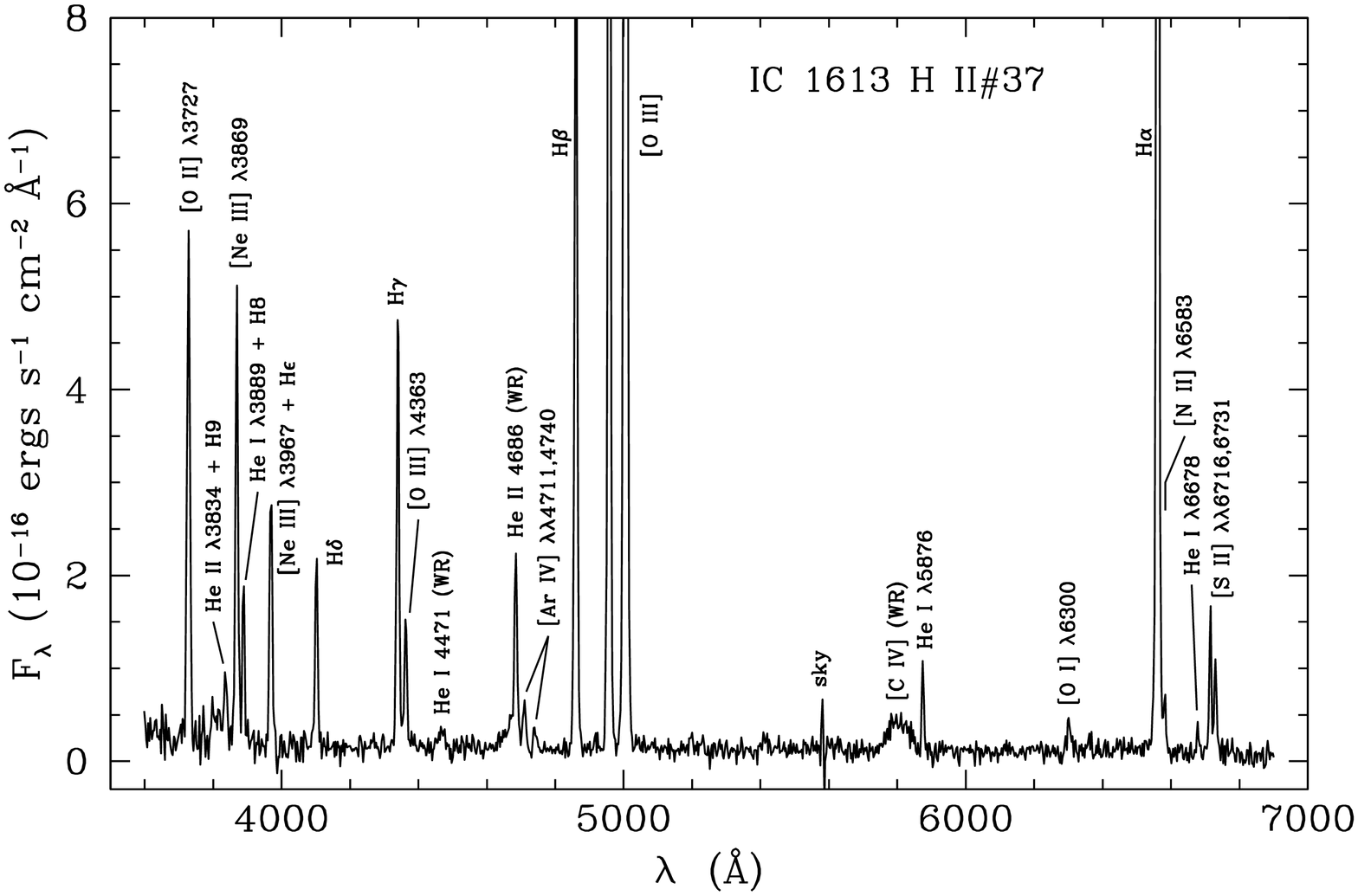}  
\caption{
IC~1613 \ion{H}{II}\#37: The spectrum is redisplayed 
to highlight faint emission lines and broad Wolf-Rayet (W-R) features.
The forbidden line [Ar~IV]$\lambda$4711 is likely blended with
He~I$\lambda$4713.
}
\label{fig_wr1613}
\end{figure*}

Data reductions were carried out in the standard manner using 
IRAF\footnote{
IRAF is distributed by the National Optical 
Astronomical Observatories, which is operated by the Associated
Universities for Research in Astronomy, Inc., under contract to the
National Science Foundation.}
routines.
The raw two-dimensional images were subtracted for bias and trimmed.
Dome flat exposures were used to remove pixel-to-pixel variations 
in response. 
Twilight flats were acquired at dusk each night to correct
for variations over larger spatial scales.
To correct for the ``slit function'' in the spatial direction, the
variation of illumination along the slit was taken into account
using dome and twilight flats. 
Multiple exposures of a given galaxy facilitated the removal of cosmic
rays. 
Wavelength calibration was obtained using helium-argon (He-Ar) arc
lamp exposures taken throughout each night.
Flux calibration was obtained using exposures of standard stars
Feige~56, LTT~1788, LTT~3864 (February run); 
and LTT~1020, LTT~7379, LTT~9239 (August run). 
The flux accuracy is listed in Table~\ref{table_obslog}.
Final one-dimensional spectra for each \ion{H}{II} region were obtained via
unweighted summed extractions.  
For \ion{H}{II} regions which were previously unidentified 
(i.e., other than those listed in Col.~1, Table~\ref{table_obslog}),
individual \ion{H}{II} regions or aperture extractions are numbered
along the slit increasing to the east.
Representative \ion{H}{II} region spectra are shown in
Fig.~\ref{fig_allspc}; the spectrum for IC~1613 \ion{H}{II} region
\#37 is displayed in Fig.~\ref{fig_wr1613} to highlight faint emission
lines.

\section{Measurements and Analysis}
\label{sec_analysis}

Emission-line strengths were measured using locally-developed
software.
Flux ratios were corrected for underlying Balmer absorption
with an equivalent width 2~\AA\ \citep{mrs85}.
Corrections and analyses were performed with SNAP
(Spreadsheet Nebular Analysis Package, \citealp{snap97}). 
In the absence of \othreea, the electron temperature was
assumed to be $T_e = 10^4$~K for computations.
Where the density-dependent line ratio, $I$(\stwoa)/$I$(\stwob),
could not be formed, a value of $n_e$ = 100~cm$^{-3}$ was adopted.
Values of the reddening are derived using \halpha\ and \hbeta\
fluxes with the method described in \cite{lee03}.
Errors in the reddening are computed from the maximum and minimum
values of the reddening based upon $2\sigma$ errors in
fits to emission lines. 

Observed flux $(F)$ and corrected intensity $(I)$ ratios are listed
in Tables~\ref{table_spec_data1} to \ref{table_spec_data4} inclusive.
The listed errors for the observed flux ratios at each wavelength
$\lambda$ account for the errors in the fits to the line profiles,
their surrounding continua, and the relative error in the sensitivity
function stated in Table~\ref{table_obslog}.  
Errors for observed ratios do not include the error in the flux
at the \hbeta\ reference line.
The uncertainty in the correction for underlying Balmer absorption
was assumed to be zero.
Errors in the corrected intensity ratios account for maximum and
minimum errors in the flux of the specified line and of 
the \hbeta\ reference line; errors in the reddening are 
not included.
\begin{table}
\scriptsize
\begin{center}
\renewcommand{\arraystretch}{1.}
\caption{
Observed and corrected line ratios for Centaurus A group dwarf galaxies.
Wavelengths are listed in \AA.
For a given galaxy whose \ion{H}{II} regions have not been previously
identified (see Table~\ref{table_obslog}), the number of the
\ion{H}{II} region aperture extraction increases to the east.
$F$ is the observed flux ratio with respect to \hbeta.
$I$ is the corrected intensity ratio, corrected for underlying
Balmer absorption and the adopted reddening listed
in Table~\ref{table_derivedprops}.
The errors in the observed line ratios account for the errors in
the fits to the line profiles, the surrounding continua, and
the relative error in the sensitivity function listed in
Table~\ref{table_obslog}; flux errors in the 
\hbeta\ reference line are not included.
Errors in the corrected line ratios account for errors in the
specified line and in the \hbeta\ reference line.
%
\label{table_spec_data1}
}
\begin{tabular}{lcccccc}
\hline \hline
& \multicolumn{2}{c}{A1243$-$335 \ion{H}{II}\#A} & 
\multicolumn{2}{c}{A1324$-$412 ap1} &
\multicolumn{2}{c}{A1334$-$277 ap1} \\
\cline{2-7}
\multicolumn{1}{l}{Identification (\AA)} &
\multicolumn{1}{c}{$F$} & \multicolumn{1}{c}{$I$} &
\multicolumn{1}{c}{$F$} & \multicolumn{1}{c}{$I$} &
\multicolumn{1}{c}{$F$} & \multicolumn{1}{c}{$I$} \\ 
\hline
$[{\rm O\;II}]\;3727$ &
        $107 \pm 15$ & $117 \pm 31$ & 
        \ldots & \ldots & 
        $105 \pm 15$ & $94 \pm 31$
\\
${\rm H}\beta\;4861$ &
        $100 \pm 10$ & $100 \pm 12$ &
        $100 \pm 15$ & $100 \pm 19$ &
        $100 \pm 13$ & $100 \pm 17$ 
\\
$[{\rm O\;III}]\;4959$ &
        $163.0 \pm 7.1$ & $145 \pm 26$ &
        \ldots & \ldots &
        $70 \pm 12$ & $63 \pm 22$
\\ 
$[{\rm O\;III}]\;5007$ &
        $456.6 \pm 8.5$ & $405 \pm 65$ &
        \ldots & \ldots &
        $176 \pm 14$ & $158 \pm 43$ 
\\
${\rm H}\alpha\;6563$ &
	$357.4 \pm 7.5$ & $286 \pm 47$ &
	$366 \pm 20$ & $286 \pm 80$ &
	$291 \pm 12$ & $268 \pm 66$
\\
$[{\rm N\;II}]\;6583$ &
	$4.7 \pm 6.1$ & $3.7 \pm 5.3$ & 
	\ldots & \ldots &
	$12 \pm 10$ & $10 \pm 11$
\\
\hline
& \multicolumn{2}{c}{A1346$-$358 \ion{H}{II}\#A} & 
\multicolumn{2}{c}{DDO 161 ap1} &
\multicolumn{2}{c}{NGC 5264 ap1} \\
\cline{2-7}
\multicolumn{1}{l}{Identification (\AA)} &
\multicolumn{1}{c}{$F$} & \multicolumn{1}{c}{$I$} &
\multicolumn{1}{c}{$F$} & \multicolumn{1}{c}{$I$} &
\multicolumn{1}{c}{$F$} & \multicolumn{1}{c}{$I$} \\ 
\hline
$[{\rm O\;II}]\;3727$ &
         $235.5 \pm 1.9$ & $415 \pm 36$ & 
         $190.5 \pm 3.6$ & $315 \pm 29$ & 
         $377 \pm 20$ & $390 \pm 61$
\\
He II 3834 $+$ H9 &
         $5.3 \pm 1.0$ & $13.9 \pm 4.5$ &
         \ldots & \ldots &
	 \ldots & \ldots
\\
$[{\rm Ne\;III}]\;3869$ &
         $19.8 \pm 1.2$ & $31.7 \pm 3.7$ & 
         \ldots & \ldots & 
         \ldots & \ldots
\\
He I 3889 $+$ H8 &
         $18.3 \pm 1.2$ & $33.7 \pm 4.4$ &
         \ldots & \ldots &
	 \ldots & \ldots
\\
$[{\rm Ne\;III}]\;3967 + {\rm H}\epsilon$ &
         $19.05 \pm 0.77$ & $32.8 \pm 3.5$ &
         \ldots & \ldots &
	 \ldots & \ldots
\\
${\rm H}\delta\;4101$ &
         $20.29 \pm 0.79$ & $31.7 \pm 3.3$ &
         $14.4 \pm 1.9$ & $26.7 \pm 6.2$ &
         \ldots & \ldots
\\
${\rm H}\gamma\;4340$ &
         $47.14 \pm 0.75$ & $60.6 \pm 5.4$ & 
         $46.4 \pm 2.4$ & $61.4 \pm 7.3$ & 
         $27.3 \pm 8.3$ & $43 \pm 25$
\\
$[{\rm O\;III}]\;4363$ &
         $1.93 \pm 0.59$ & $2.34 \pm 0.82$ &
         $< 6.6 $ & $< 7.7$ & 
         \ldots & \ldots
\\
He I 4472 &
        $3.52 \pm 0.53$ & $4.04 \pm 0.81$ &
        \ldots & \ldots &
	\ldots & \ldots
\\
${\rm H}\beta\;4861$ &
        $100.0 \pm 1.4$ & $100.0 \pm 4.4$ &
        $100.0 \pm 2.1$ & $100.0 \pm 4.8$ &
        $100.0 \pm 7.2$ & $100.0 \pm 9.3$ 
\\
$[{\rm O\;III}]\;4959$ &
        $95.6 \pm 2.3$ & $90.6 \pm 8.3$ &
        $92.8 \pm 2.3$ & $85.9 \pm 8.2$ &
        $34.4 \pm 5.7$ & $28.8 \pm 7.6$
\\ 
$[{\rm O\;III}]\;5007$ &
        $292.8 \pm 2.8$ & $273 \pm 24$ &
        $278.9 \pm 2.8$ & $254 \pm 23$ &
        $83.4 \pm 7.1$ & $69 \pm 13$
\\
He I 5876 & 
	$15.46 \pm 0.50$ & $11.4 \pm 1.1$ &
	$17.0 \pm 1.9$ & $12.4 \pm 2.1$ &
	\ldots & \ldots
\\
$[{\rm N\;II}]\;6548$ &
	$15.3 \pm 2.7$ & $9.9 \pm 2.2$ & 
	\ldots & \ldots &
	$20.4 \pm 7.6$ & $15.0 \pm 7.1$
\\
${\rm H}\alpha\;6563$ &
	$443.0 \pm 3.4$ & $286 \pm 25$ &
	$440.2 \pm 3.0$ & $286 \pm 26$ &
	$374.6 \pm 9.5$ & $286 \pm 39$
\\
$[{\rm N\;II}]\;6583$ &
	$39.6 \pm 2.7$ & $25.4 \pm 3.2$ &
	$8.9 \pm 2.5$ & $5.7 \pm 1.9$ &
	$63.6 \pm 7.8$ & $47 \pm 10$
\\
He I 6678 &
        $4.04 \pm 0.68$ & $2.55 \pm 0.56$ &
        \ldots & \ldots &
	\ldots & \ldots
\\
$[{\rm S\;II}]\;6716$ &
	$39.45 \pm 0.80$ & $24.8 \pm 2.2$ &
	$22.0 \pm 2.0$ & $13.9 \pm 2.0$ &
	$93.4 \pm 7.7$ & $68 \pm 12$
\\
$[{\rm S\;II}]\;6731$ &
	$27.45 \pm 0.75$ & $17.2 \pm 1.4$ &
	$15.8 \pm 1.9$ & $10.0 \pm 1.7$ &
	$60.5 \pm 7.2$ & $44.0 \pm 9.6$
\\ 
\hline
\end{tabular}
\end{center}
\end{table}

\begin{table}
\scriptsize
\begin{center}
\renewcommand{\arraystretch}{1.}
\caption{
Observed and corrected line ratios for Sculptor group dwarf galaxies;
same comments as Table~\ref{table_spec_data1}.
NOTE:
$^a$~The recession velocity of emission lines is 
$v_{\odot}/c \simeq +0.11$, placing this galaxy in the background. 
No corrections are applied.
%
\label{table_spec_data2}
}
\begin{tabular}{lcccccc}
\hline \hline
& \multicolumn{2}{c}{AM0106$-$382 ap1} & 
\multicolumn{2}{c}{AM0106$-$382 ap2} &
\multicolumn{2}{c}{AM0106$-$382 ap3} \\
\cline{2-7}
\multicolumn{1}{l}{Identification (\AA)} &
\multicolumn{1}{c}{$F$} & \multicolumn{1}{c}{$I$} &
\multicolumn{1}{c}{$F$} & \multicolumn{1}{c}{$I$} &
\multicolumn{1}{c}{$F$} & \multicolumn{1}{c}{$I$} \\ 
\hline
$[{\rm O\;II}]\;3727$ &
        $168 \pm 21$ & $155 \pm 40$ & 
        $165 \pm 15$ & $149 \pm 29$ & 
        $178 \pm 17 $ & $158 \pm 31$
\\
${\rm H}\beta\;4861$ &
        $100.0 \pm 9.9$ & $100.0 \pm 12$ &
        $100.0 \pm 6.4$ & $100.0 \pm 8.9$ &
        $100.0 \pm 6.4$ & $100.0 \pm 8.9$ 
\\
$[{\rm O\;III}]\;4959$ &
        $22.3 \pm 7.7$ & $20.6 \pm 9.8$ &
        $37.6 \pm 4.7$ & $33.9 \pm 7.6$ &
        $45.8 \pm 5.7$ & $40.6 \pm 9.1$
\\ 
$[{\rm O\;III}]\;5007$ &
        $62.4 \pm 7.9$ & $58 \pm 15$ &
        $124.8 \pm 5.8$ & $113 \pm 18$ &
        $162.1 \pm 7.1$ & $144 \pm 22$ 
\\
$[{\rm N\;II}]\;6548$ &
	\ldots & \ldots & 
	$1.3 \pm 4.2$ & $1.1 \pm 3.9$ &
	$2.8 \pm 3.9$ & $2.5 \pm 3.7$
\\
${\rm H}\alpha\;6563$ &
	$239.7 \pm 8.9$ & $225 \pm 42$ &
	$266.2 \pm 5.3$ & $245 \pm 35$ &
	$287.4 \pm 4.9$ & $260 \pm 37$
\\
$[{\rm N\;II}]\;6583$ &
	$12.4 \pm 7.2$ & $11.4 \pm 8.1$ & 
	$7.0 \pm 4.3$ & $6.3 \pm 4.4$ &
	$5.02 \pm 3.9$ & $4.4 \pm 3.9$ 
\\
$[{\rm S\;II}]\;6716$ &
	$39.8 \pm 8.0$ & $37 \pm 12$ &
	$29.1 \pm 5.2$ & $26.3 \pm 7.2$ &
	$23.3 \pm 4.1$ & $20.6 \pm 5.6$
\\
$[{\rm S\;II}]\;6731$ &
	$20.7 \pm 7.0$ & $19.1 \pm 8.9$ &
	$15.2 \pm 4.4$ & $13.7 \pm 5.2$ &
	$15.7 \pm 3.3$ & $13.9 \pm 4.3$
\\
\hline
& \multicolumn{2}{c}{ESO 347$-$G017 ap1} & 
\multicolumn{2}{c}{ESO 347$-$G017 ap2} &
\multicolumn{2}{c}{ESO 347$-$G017 ap3} \\
\cline{2-7}
\multicolumn{1}{l}{Identification (\AA)} &
\multicolumn{1}{c}{$F$} & \multicolumn{1}{c}{$I$} &
\multicolumn{1}{c}{$F$} & \multicolumn{1}{c}{$I$} &
\multicolumn{1}{c}{$F$} & \multicolumn{1}{c}{$I$} \\ 
\hline
$[{\rm O\;II}]\;3727$ &
         $215.3 \pm 3.4$ & $200 \pm 22$ & 
         $159.7 \pm 4.2$ & $171 \pm 21$ & 
         $244 \pm 21$ & $296 \pm 69$
\\
$[{\rm Ne\;III}]\;3869$ &
         $39.0 \pm 4.3$ & $36.1 \pm 6.5$ & 
         $44.2 \pm 6.4$ & $45.5 \pm 9.9$ & 
         \ldots & \ldots
\\
${\rm H}\gamma\;4340$ &
         $33.5 \pm 1.9$ & $42.7 \pm 6.5$ & 
         $28.0 \pm 3.1$ & $41.7 \pm 9.3$ & 
         \ldots & \ldots
\\
$[{\rm O\;III}]\;4363$ &
         $<$ 4.9 (2$\sigma$) & $<$ 4.4 (2$\sigma$) &
         $<$ 4.2 (2$\sigma$) & $<$ 4.4 (2$\sigma$) & 
         \ldots & \ldots
\\
${\rm H}\beta\;4861$ &
        $100.0 \pm 2.3$ & $100.0 \pm 5.8$ &
        $100.0 \pm 3.5$ & $100.0 \pm 6.5$ &
        $100 \pm 10$ & $100 \pm 13$ 
\\
$[{\rm O\;III}]\;4959$ &
         $109.2 \pm 3.2$ & $98 \pm 11$ &
         $136.0 \pm 3.1$ & $118 \pm 14$ &
         $72.6 \pm 6.8$ & $62 \pm 14$
\\ 
$[{\rm O\;III}]\;5007$ &
         $312.4 \pm 3.9$ & $281 \pm 31$ &
         $414.0 \pm 3.8$ & $359 \pm 42$ &
         $192.5 \pm 8.4$ & $160 \pm 31$
\\
${\rm He\;I}\;5876$ &
	 $17.2 \pm 2.7$ & $15.4 \pm 3.4$ &
	 $20.3 \pm 2.4$ & $16.3 \pm 3.1$ &
	 \ldots & \ldots
\\
$[{\rm N\;II}]\;6548$ &
	 $8.7 \pm 2.8$ & $7.7 \pm 2.9$ & 
	 $7.0 \pm 2.7$ & $5.6 \pm 2.5$ &
	 \ldots & \ldots
\\
${\rm H}\alpha\;6563$ &
	 $315.9 \pm 3.5$ & $286 \pm 31$ &
	 $363.6 \pm 3.4$ & $286 \pm 33$ &
	 $418 \pm 12$ & $286 \pm 55$
\\
$[{\rm N\;II}]\;6583$ &
	 $11.8 \pm 2.9$ & $10.5 \pm 3.2$ &
	 $12.2 \pm 2.8$ & $9.6 \pm 2.8$ &
	 $10.8 \pm 9.8$ & $5.8 \pm 7.7$
\\
$[{\rm S\;II}]\;6716$ &
	 $42.0 \pm 1.8$ & $37.0 \pm 4.6$ &
	 $24.6 \pm 2.0$ & $18.7 \pm 3.0$ &
	 $41.9 \pm 7.9$ & $27.8 \pm 9.1$
\\
$[{\rm S\;II}]\;6731$ &
	 $25.3 \pm 1.6$ & $22.3 \pm 3.1$ &
	 $17.3 \pm 1.9$ & $13.1 \pm 2.4$ &
	 $20.7 \pm 6.3$ & $13.7 \pm 6.1$
\\ 
\hline
& \multicolumn{2}{c}{ESO 348$-$G009 ap1} & 
\multicolumn{2}{c}{ESO 348$-$G009 ap2$\,^a$ } &
\multicolumn{2}{c}{UGCA 442 \ion{H}{II}\#2} \\
\cline{2-7}
\multicolumn{1}{l}{Identification (\AA)} &
\multicolumn{1}{c}{$F$} & \multicolumn{1}{c}{$I$} &
\multicolumn{1}{c}{$F$} & \multicolumn{1}{c}{$I$} &
\multicolumn{1}{c}{$F$} & \multicolumn{1}{c}{$I$} \\ 
\hline
$[\rm{O\;II}]\;3727$ &
        $317 \pm 29$ & $283 \pm 72$ & 
        $589 \pm 54$ & \ldots  & 
        $221.4 \pm 4.9$ & $237 \pm 27$
\\
${\rm H}\gamma\;4340$ &
        \ldots & \ldots & 
        \ldots & \ldots & 
        $36.2 \pm 2.4$ & $47.3 \pm 7.6$
\\
$[{\rm O\;III}]\;4363$ &
        \ldots & \ldots & 
        \ldots & \ldots & 
        $<$ 3.8 (2$\sigma$) & $<$ 3.6 (2$\sigma$)
\\
${\rm H}\beta\;4861$ &
        $100 \pm 12$ & $100 \pm 15$ &
        $100 \pm 33$ & \ldots &
        $100.0 \pm 2.5$ & $100.0 \pm 5.9$ 
\\
$[{\rm O\;III}]\;4959$ &
        $53 \pm 10$ & $47 \pm 16$ &
        $101 \pm 32$ & \ldots &
        $53.6 \pm 2.2$ & $47.5 \pm 5.9$
\\ 
$[{\rm O\;III}]\;5007$ &
        $118 \pm 12$ & $106 \pm 28$ &
        $583 \pm 41$ & \ldots &
        $155.0 \pm 2.7$ & $137 \pm 15$
\\
${\rm He\;I}\;5876$ &
	\ldots & \ldots &
	\ldots & \ldots &
	$16.5 \pm 2.8$ & $13.5 \pm 3.2$
\\
$[{\rm N\;II}]\;6548$ &
	\ldots & \ldots & 
	\ldots & \ldots &
	$6.3 \pm 2.3$ & $4.9 \pm 2.2$
\\
${\rm H}\alpha\;6563$ &
	$254.1 \pm 9.1$ & $233 \pm 49$ &
	\ldots & \ldots &
	$356.2 \pm 2.9$ & $286 \pm 32$
\\
$[{\rm N\;II}]\;6583$ &
	$10.7 \pm 7.3$ & $9.6 \pm 8.0$ &
	\ldots & \ldots &
	$14.3 \pm 2.4$ & $11.3 \pm 2.6$
\\
$[{\rm S\;II}]\;6716$ &
	$44.2 \pm 7.7$ & $39 \pm 13$ &
	\ldots & \ldots &
	$36.8 \pm 1.5$ & $28.7 \pm 3.6$
\\
$[{\rm S\;II}]\;6731$ &
	$20.6 \pm 5.9$ & $18.4 \pm 8.1$ &
	\ldots & \ldots &
	$22.7 \pm 1.4$ & $17.7 \pm 2.5$
\\ 
\hline
\end{tabular}
\end{center}
\end{table}

\begin{table}
\scriptsize
\begin{center}
\renewcommand{\arraystretch}{1.}
\caption{
Observed and corrected line ratios for other southern dwarf galaxies;
same comments as Table~\ref{table_spec_data1}.
ESO302$-$G014 \#1 is not included here, as only \othreec\ and \halpha\
were detected; the measured fluxes in ergs~s$^{-1}$~cm$^{-2}$ are 
$(2.13 \pm 0.44) \times 10^{-15}$ and 
$(2.73 \pm 0.43) \times 10^{-15}$, respectively.
NOTE:
$^a$~A broad Wolf-Rayet feature was detected near He~I$\lambda$4472.
%
\label{table_spec_data3}
}
\begin{tabular}{lcccccc}
\hline \hline
& \multicolumn{2}{c}{A0355$-$465 \ion{H}{II}\#B} & 
\multicolumn{2}{c}{ESO358$-$G060 ap1} &
\multicolumn{2}{c}{IC 1613 \ion{H}{II}\#13} \\
\cline{2-7}
\multicolumn{1}{l}{Identification (\AA)} &
\multicolumn{1}{c}{$F$} & \multicolumn{1}{c}{$I$} &
\multicolumn{1}{c}{$F$} & \multicolumn{1}{c}{$I$} & 
\multicolumn{1}{c}{$F$} & \multicolumn{1}{c}{$I$} \\ 
\hline
$[{\rm O\;II}]\;3727$ &
        $185.9 \pm 5.3$ & $277 \pm 28$ &
        $66.1 \pm 4.9$ & $64 \pm 12$ &
        $256.7 \pm 6.1$ & $251 \pm 28$ 
\\
$[{\rm Ne\;III}]\;3869$ &
        $41.8 \pm 7.0$ & $58 \pm 13$ &
        $7.2 \pm 2.1$ & $7.0 \pm 2.6$ &
        \ldots & \ldots 
\\
He I 3889 $+$ H8 &
        $16.7 \pm 5.6$ & $31 \pm 16$ &
        $8.5 \pm 2.7$ & $11.5 \pm 6.0$ &
        \ldots & \ldots 
\\
$[{\rm Ne\;III}]\;3967 + {\rm H}\epsilon$ &
        $19.1 \pm 4.9$ & $32 \pm 12$ &
        $9.7 \pm 2.7$ & $12.8 \pm 6.0$ &
        $19.08 \pm 4.1$ & $22.9 \pm 7.5$ 
\\
${\rm H}\delta\;4101$ &
        $29.8 \pm 4.5$ & $43 \pm 10$ &
        $20.5 \pm 3.8$ & $23.2 \pm 7.2$ &
        $20.9 \pm 3.4$ & $24.0 \pm 6.1$ 
\\
${\rm H}\gamma\;4340$ &
        $45.5 \pm 4.5$ & $57.1 \pm 9.6$ &
        $42.8 \pm 3.4$ & $45.0 \pm 8.7$ &
        $38.8 \pm 3.1$ & $41.1 \pm 6.5$
\\
$[{\rm O\;III}]\;4363$ &
        $<$ 6.4 (2$\sigma$) & $<$ 7.1 (2$\sigma$) &
        $<$ 5.8 (2$\sigma$) & $<$ 5.6 (2$\sigma$) &
        $<$ 7.3 (2$\sigma$) & $<$ 7.1 (2$\sigma$)
\\
${\rm H}\beta\;4861$ &
        $100.0 \pm 2.8$ & $100.0 \pm 5.1$ &
        $100.0 \pm 4.1$ & $100.0 \pm 8.1$ &
        $100.0 \pm 2.3$ & $100.0 \pm 5.7$ 
\\
$[{\rm O\;III}]\;4959$ &
        $166.8 \pm 5.2$ & $156 \pm 16$ &
        $68.7 \pm 2.8$ & $66 \pm 11$ &
        $67.1 \pm 1.6$ & $65.6 \pm 7.3$ 
\\ 
$[{\rm O\;III}]\;5007$ &
        $486.1 \pm 6.4$ & $450 \pm 43$ &
        $202.4 \pm 3.4$ & $196 \pm 29$ &
        $204.4 \pm 1.9$ & $200 \pm 22$ 
\\
${\rm He\;I}\;5876$ &
        \ldots & \ldots &
        \ldots & \ldots &
	$11.4 \pm 4.3$ & $11.2 \pm 4.9$ 
\\
$[{\rm N\;II}]\;6548$ &
        \ldots & \ldots &
        \ldots & \ldots &
	$7.3 \pm 1.7$ & $7.2 \pm 2.1$ 
\\
${\rm H}\alpha\;6563$ &
        $403.6 \pm 3.7$ & $286 \pm 27$ &
        $258.2 \pm 3.2$ & $251 \pm 37$ &
	$227.5 \pm 2.1$ & $223 \pm 24$ 
\\
$[{\rm N\;II}]\;6583$ &
        $15.6 \pm 3.0$ & $10.9 \pm 2.7$ &
        $4.5 \pm 2.6$ & $4.4 \pm 2.9$ &
	$10.9 \pm 1.7$ & $10.7 \pm 2.4$
\\
$[{\rm S\;II}]\;6716$ &
        $32.6 \pm 3.5$ & $22.5 \pm 3.8$ &
        $12.3 \pm 2.1$ & $11.9 \pm 3.2$ &
	$19.7 \pm 1.8$ & $19.3 \pm 3.1$ 
\\
$[{\rm S\;II}]\;6731$ &
        $23.1 \pm 3.3$ & $15.9 \pm 3.2$ &
        $6.0 \pm 1.7$ & $5.8 \pm 2.2$ &
	$11.2 \pm 1.6$ & $10.9 \pm 2.3$ 
\\
\hline
& \multicolumn{2}{c}{IC 1613 \ion{H}{II}\#37} & 
\multicolumn{2}{c}{IC 2032 ap1} &
\multicolumn{2}{c}{IC 5152 \ion{H}{II}\#A} \\
\cline{2-7}
\multicolumn{1}{l}{Identification (\AA)} &
\multicolumn{1}{c}{$F$} & \multicolumn{1}{c}{$I$} &
\multicolumn{1}{c}{$F$} & \multicolumn{1}{c}{$I$} &
\multicolumn{1}{c}{$F$} & \multicolumn{1}{c}{$I$} \\ 
\hline
$[{\rm O\;II}]\;3727$ &
        $61.0 \pm 1.6$ & $70.1 \pm 7.7$ &
        $320 \pm 28$ & $286 \pm 59$ &
	$220.0 \pm 2.5$ & $211 \pm 22$ 
\\
He II 3834 $+$ H9 &
        $6.3 \pm 1.1$ & $7.9 \pm 2.1$ &
        \ldots & \ldots &
        \ldots & \ldots 
\\
$[{\rm Ne\;III}]\;3869$ &
        $47.7 \pm 1.4$ & $53.5 \pm 6.0$ &
        $55 \pm 14$ & $49 \pm 18$ &
        $21.3 \pm 2.7$ & $20.3 \pm 3.8$ 
\\
He I 3889 $+$ H8 &
        $15.6 \pm 1.2$ & $18.0 \pm 2.7$ &
        \ldots & \ldots &
        $12.5 \pm 2.1$ & $16.9 \pm 4.7$ 
\\
$[{\rm Ne\;III}]\;3967 + {\rm H}\epsilon$ &
        $31.6 \pm 1.1$ & $35.2 \pm 4.1$ &
        \ldots & \ldots &
        $11.7 \pm 2.0$ & $16.4 \pm 4.7$ 
\\
${\rm H}\delta\;4101$ &
        $20.16 \pm 0.89$ & $22.3 \pm 2.7$ &
        \ldots & \ldots &
	$18.3 \pm 1.8$ & $22.5 \pm 4.2$
\\
${\rm H}\gamma\;4340$ &
        $47.52 \pm 0.84$ & $50.4 \pm 5.4$ &
        \ldots & \ldots &
        $42.1 \pm 1.3$ & $45.1 \pm 5.2$
\\
$[{\rm O\;III}]\;4363$ &
        $13.83 \pm 0.70$ & $14.5 \pm 1.8$ &
        \ldots & \ldots &
        $3.6 \pm 1.1$ & $3.4 \pm 1.2$ 
\\
He I 4472 &
        \ldots$\,^a$ & \ldots & 
        \ldots & \ldots &
	$5.98 \pm 0.89$ & $5.7 \pm 1.2$ 
\\
He II 4686 &
	$21.45 \pm 0.85$ & $21.7 \pm 2.5$ &
	\ldots & \ldots &
	\ldots & \ldots 
\\
He I 4713 & 
	$4.31 \pm 0.68$ & $4.35 \pm 0.96$ &
	\ldots & \ldots &
	\ldots & \ldots
\\
$[{\rm Ar\;IV}]\;4740$ &
        $2.01 \pm 0.68$ & $2.02 \pm 0.80$ &
        \ldots & \ldots &
        \ldots & \ldots 
\\
${\rm H}\beta\;4861$ &
        $100.0 \pm 1.6$ & $100.0 \pm 5.3$ &
        $100.0 \pm 9.4$ & $100 \pm 11$ &
        $100.0 \pm 1.3$ & $100 \pm 5.3$ 
\\
$[{\rm O\;III}]\;4959$ &
        $176.2 \pm 4.4$ & $174 \pm 19$ &
        $79.0 \pm 7.6$ & $70 \pm 15$ &
        $98.1 \pm 3.4$ & $94 \pm 11$ 
\\ 
$[{\rm O\;III}]\;5007$ &
        $531.6 \pm 5.5$ & $523 \pm 55$ &
        $213.9 \pm 9.3$ & $189 \pm 32$ &
        $294.3 \pm 4.2$ & $282 \pm 30$ 
\\
${\rm He\;I}\;5876$ &
        $6.94 \pm 0.68$ & $6.4 \pm 1.1$ &
        \ldots & \ldots &
	$12.7 \pm 1.3$ & $12.1 \pm 2.0$ 
\\
$[{\rm O\;I}]\;6300$ &
        $4.35 \pm 0.77$ & $3.95 \pm 0.94$ &
        \ldots & \ldots &
        $5.59 \pm 0.76$ & $5.4 \pm 1.1$ 
\\
$[{\rm S\;III}]\;6312$ &
        $1.11 \pm 0.67$ & $1.01 \pm 0.67$ &
        \ldots & \ldots &
        \ldots & \ldots 
\\
$[{\rm O\;I}]\;6363$ &
        $1.78 \pm 0.59$ & $1.62 \pm 0.63$ &
        \ldots & \ldots &
        $1.38 \pm 0.60$ & $1.32 \pm 0.65$ 
\\
$[{\rm N\;II}]\;6548$ &
        $4.6 \pm 2.2$ & $4.1 \pm 2.3$ &
        \ldots & \ldots &
	$11.4 \pm 2.8$ & $10.9 \pm 3.3$ 
\\
${\rm H}\alpha\;6563$ &
        $307.0 \pm 2.8$ & $277 \pm 29$ &
        $316.6 \pm 8.94$ & $286 \pm 46$ &
	$270.8 \pm 3.5$ & $262 \pm 27$ 
\\
$[{\rm N\;II}]\;6583$ &
        $4.7 \pm 2.3$ & $4.2 \pm 2.3$ &
        $19.8 \pm 7.2$ & $17.4 \pm 8.4$ &
	$20.8 \pm 2.9$ & $20.0 \pm 4.0$ 
\\
${\rm He\;I}\;6678$ &
        $2.29 \pm 0.46$ & $2.05 \pm 0.53$ &
        \ldots & \ldots &
	$4.54 \pm 0.53$ & $4.34 \pm 0.78$ 
\\
$[{\rm S\;II}]\;6716$ &
        $12.93 \pm 0.54$ & $11.6 \pm 1.4$ &
        $48.3 \pm 8.3$ & $42 \pm 12$ &
	$19.93 \pm 0.59$ & $19.1 \pm 2.1$ 
\\
$[{\rm S\;II}]\;6731$ &
        $8.45 \pm 0.50$ & $7.55 \pm 0.99$ &
        $42.9 \pm 8.3$ & $38 \pm 12$ &
	$14.35 \pm 0.56$ & $13.7 \pm 1.6$ 
\\ 
\hline
\end{tabular}
\end{center}
\end{table}

\begin{table}
\scriptsize
\begin{center}
\renewcommand{\arraystretch}{1.}
\caption{
Observed and corrected line ratios for other southern dwarf galaxies;
same comments as Table~\ref{table_spec_data1}.
%
\label{table_spec_data4}
}
\begin{tabular}{lcccccccc}
\hline \hline
& \multicolumn{2}{c}{NGC 2915 ap1} &
\multicolumn{2}{c}{NGC 2915 ap2} & 
\multicolumn{2}{c}{NGC 3109 \ion{H}{II}\#6 ap1} &
\multicolumn{2}{c}{NGC 3109 \ion{H}{II}\#6 ap2} \\
\cline{2-9}
\multicolumn{1}{l}{Identification (\AA)} &
\multicolumn{1}{c}{$F$} & \multicolumn{1}{c}{$I$} &
\multicolumn{1}{c}{$F$} & \multicolumn{1}{c}{$I$} &
\multicolumn{1}{c}{$F$} & \multicolumn{1}{c}{$I$} &
\multicolumn{1}{c}{$F$} & \multicolumn{1}{c}{$I$} \\ 
\hline
$[{\rm O\;II}]\;3727$ &
        $396 \pm 15$ & $660 \pm 110$ & 
        $306 \pm 35$ & $440 \pm 110$ &
        $178 \pm 26$ & $152 \pm 60$ &
        $65 \pm 16$ & $51 \pm 25$ 
\\
${\rm H}\beta\;4861$ &
        $100.0 \pm 7.5$ & $100 \pm 12$ & 
        $100.0 \pm 9.4$ & $100 \pm 12$ &
        $100 \pm 18$ & $100 \pm 23$ &
        $100 \pm 17$ & $100 \pm 23$ 
\\
$[{\rm O\;III}]\;4959$ &
        $198.9 \pm 7.7$ & $115 \pm 20$ & 
        $63 \pm 10$ & $54 \pm 15$ &
        $36 \pm 18$ & $30 \pm 23$ &
        $91 \pm 11$ & $72 \pm 26$ 
\\ 
$[{\rm O\;III}]\;5007$ &
        $549.6 \pm 9.2$ & $310 \pm 50$ & 
        $216 \pm 13$ & $181 \pm 34$ &
        $166 \pm 24$ & $142 \pm 56$ &
        $200 \pm 13$ & $157 \pm 50$ 
\\
$[{\rm N\;II}]\;6548$ &
        $22.1 \pm 6.5$ & $6.7 \pm 2.8$ & 
        $11.4 \pm 8.2$ & $7.0 \pm 5.9$ &
        \ldots & \ldots &
        \ldots & \ldots 
\\
${\rm H}\alpha\;6563$ &
        $886.6 \pm 8.1$ & $286 \pm 45$ & 
        $447 \pm 10$ & $286 \pm 47$ &
        $264 \pm 18$ & $237 \pm 78$ &
        $300 \pm 17$ & $248 \pm 78$ 
\\
$[{\rm N\;II}]\;6583$ &
        $74.8 \pm 6.6$ & $22.4 \pm 4.8$ & 
        $64.2 \pm 8.4$ & $39 \pm 10$ &
        \ldots & \ldots &
        \ldots & \ldots 
\\
$[{\rm S\;II}]\;6716$ &
        $101.7 \pm 5.5$ & $29.4 \pm 5.4$ & 
        $73.6 \pm 7.6$ & $44 \pm 10$ &
        \ldots & \ldots &
        \ldots & \ldots 
\\
$[{\rm S\;II}]\;6731$ &
        $70.2 \pm 5.3$ & $20.2 \pm 4.1$ & 
        $32.5 \pm 6.8$ & $19.4 \pm 6.4$ &
        \ldots & \ldots &
        \ldots & \ldots 
\\
\hline
& \multicolumn{2}{c}{NGC 3109 \ion{H}{II}\#6 ap3} & 
\multicolumn{2}{c}{NGC 3109 \ion{H}{II}\#6 ap4} & 
\multicolumn{2}{c}{NGC 3109 \ion{H}{II}\#6 ap5} & & \\
\cline{2-9}
\multicolumn{1}{l}{Identification (\AA)} &
\multicolumn{1}{c}{$F$} & \multicolumn{1}{c}{$I$} &
\multicolumn{1}{c}{$F$} & \multicolumn{1}{c}{$I$} &
\multicolumn{1}{c}{$F$} & \multicolumn{1}{c}{$I$} & & \\ 
\hline
$[{\rm O\;II}]\;3727$ &
        $110.5 \pm 8.6$ & $106 \pm 20$ &
        $108.9 \pm 3.7$ & $109 \pm 16$ &
        $115.6 \pm 3.8$ & $121 \pm 18$ & &
\\
$[{\rm Ne\;III}]\;3967 + {\rm H}\epsilon$ &
        \ldots & \ldots &
        $15.2 \pm 3.8$ & $18.1 \pm 7.0$ &
        $15.8 \pm 4.1$ & $19.3 \pm 7.6$ & & 
\\
${\rm H}\delta\;4101$ &
        \ldots & \ldots &
        $13.1 \pm 2.9$ & $16.8 \pm 6.2$ &
	$15.8 \pm 2.8$ & $19.7 \pm 5.9$ & & 
\\
${\rm H}\gamma\;4340$ &
        \ldots & \ldots &
        $45.8 \pm 4.0$ & $49.5 \pm 9.7$ &
        $47.3 \pm 2.9$ & $51.3 \pm 8.8$ & &
\\
${\rm H}\beta\;4861$ &
        $100.0 \pm 5.2$ & $100.0 \pm 8.9$ &
        $100.0 \pm 2.9$ & $100.0 \pm 7.5$ &
        $100.0 \pm 2.6$ & $100.0 \pm 7.3$ & &
\\
$[{\rm O\;III}]\;4959$ &
        $59.4 \pm 5.2$ & $54 \pm 11$ &
        $94.2 \pm 3.1$ & $86 \pm 13$ &
        $82.4 \pm 2.8$ & $77 \pm 11$ & &
\\ 
$[{\rm O\;III}]\;5007$ &
        $195.4 \pm 6.4$ & $178 \pm 29$ &
        $289.5 \pm 3.8$ & $265 \pm 38$ &
        $251.5 \pm 3.4$ & $234 \pm 33$ & &
\\
He I 5876 &
        $16.4 \pm 5.0$ & $14.7 \pm 5.9$ &
        $14.2 \pm 2.9$ & $12.5 \pm 3.6$ &
        $13.7 \pm 2.7$ & $12.2 \pm 3.5$ & & 
\\
$[{\rm N\;II}]\;6548$ &
        \ldots & \ldots &
        $3.4 \pm 3.1$ & $2.9 \pm 2.9$ &
        $7.2 \pm 2.7$ & $6.3 \pm 2.8$ & &
\\
${\rm H}\alpha\;6563$ &
        $318.3 \pm 5.6$ & $286 \pm 45$ &
        $320.9 \pm 3.8$ & $286 \pm 41$ &
        $323.5 \pm 3.3$ & $286 \pm 40$ & &
\\
$[{\rm N\;II}]\;6583$ &
        $15.1 \pm 4.5$ & $13.4 \pm 5.3$ &
        $14.7 \pm 3.1$ & $12.7 \pm 3.8$ &
        $16.7 \pm 2.7$ & $14.5 \pm 3.6$ & &
\\
$[{\rm S\;II}]\;6716$ &
        $23.4 \pm 2.8$ & $20.6 \pm 4.7$ &
        $22.9 \pm 1.7$ & $19.8 \pm 3.4$ &
        $22.5 \pm 1.5$ & $19.5 \pm 3.3$ & &
\\
$[{\rm S\;II}]\;6731$ &
        $14.9 \pm 2.6$ & $13.2 \pm 3.7$ &
        $13.8 \pm 1.5$ & $11.9 \pm 2.4$ &
        $14.1 \pm 1.4$ & $12.1 \pm 2.4$ & &
\\ 
\hline
& \multicolumn{2}{c}{NGC 3109 \ion{H}{II}\#6 ap6} & 
\multicolumn{2}{c}{NGC 3109 \ion{H}{II}\#6 ap7} & 
\multicolumn{2}{c}{Sag DIG \ion{H}{II}\#3} & & \\
\cline{2-9}
\multicolumn{1}{l}{Identification (\AA)} &
\multicolumn{1}{c}{$F$} & \multicolumn{1}{c}{$I$} &
\multicolumn{1}{c}{$F$} & \multicolumn{1}{c}{$I$} &
\multicolumn{1}{c}{$F$} & \multicolumn{1}{c}{$I$} & & \\ 
\hline
$[{\rm O\;II}]\;3727$ &
        $118.8 \pm 4.0$ & $129 \pm 19$ &
        $204 \pm 27$ & $270 \pm 100$ &
        $71.5 \pm 6.9$ & $95 \pm 17$ & &
\\
$[{\rm Ne\;III}]\;3967 + {\rm H}\epsilon$ &
        $14.3 \pm 3.8$ & $18.0 \pm 6.3$ &
        \ldots & \ldots &
        \ldots & \ldots & & 
\\
${\rm H}\delta\;4101$ &
        $18.9 \pm 3.2$ & $22.7 \pm 6.4$ &
        \ldots & \ldots &
	$17.5 \pm 3.3$ & $24.6 \pm 7.3$ & &
\\
${\rm H}\gamma\;4340$ &
        $46.4 \pm 3.9$ & $50.4 \pm 9.6$ &
        \ldots & \ldots &
        $48.2 \pm 4.4$ & $57 \pm 10$ & &
\\
${\rm H}\beta\;4861$ &
        $100.0 \pm 3.2$ & $100.0 \pm 7.6$ &
        $100 \pm 18$ & $100 \pm 22$ &
        $100.0 \pm 3.8$ & $100.0 \pm 6.5$ & &
\\
$[{\rm O\;III}]\;4959$ &
        $67.4 \pm 3.0$ & $65 \pm 10$ &
        $31 \pm 16$ & $28 \pm 21$ &
        $54.1 \pm 4.0$ & $51.5 \pm 8.0$ & &
\\ 
$[{\rm O\;III}]\;5007$ &
        $203.2 \pm 3.6$ & $195 \pm 28$ &
        $111 \pm 19$ & $98 \pm 40$ &
        $165.8 \pm 5.0$ & $157 \pm 19$ & &
\\
He I 5876 &
        $12.7 \pm 2.3$ & $11.7 \pm 3.2$ &
        \ldots & \ldots &
        \ldots & \ldots & & 
\\
$[{\rm N\;II}]\;6548$ &
        $11.3 \pm 2.5$ & $10.1 \pm 3.1$ &
        \ldots & \ldots &
        $4.4 \pm 3.2$ & $3.4 \pm 2.7$ & &
\\
${\rm H}\alpha\;6563$ &
        $318.3 \pm 3.2$ & $286 \pm 41$ &
        $407 \pm 17$ & $286 \pm 86$ &
        $369.1 \pm 4.1$ & $286 \pm 33$ & &
\\
$[{\rm N\;II}]\;6583$ &
        $18.4 \pm 2.6$ & $16.4 \pm 3.8$ &
        $46 \pm 14$ & $32 \pm 17$ &
        $3.5 \pm 3.3$ & $2.7 \pm 2.7$ & &
\\
$[{\rm S\;II}]\;6716$ &
        $23.5 \pm 1.6$ & $20.9 \pm 3.6$ &
        \ldots & \ldots & 
        $18.5 \pm 5.6$ & $14.1 \pm 5.3$ & & 
\\
$[{\rm S\;II}]\;6731$ &
        $14.9 \pm 1.5$ & $13.3 \pm 2.6$ &
        \ldots & \ldots & 
	$9.7 \pm 4.4$ & $7.4 \pm 3.9$ & &
\\
\hline
\end{tabular}
\end{center}
\end{table}

Derived properties are listed in Table~\ref{table_derivedprops}.
The listed properties include \hbeta\ intensities corrected for
underlying Balmer absorption and reddening, derived and adopted values
of the reddening, observed \hbeta\ emission equivalent widths, and
derived O$^+$ and O$^{+2}$ electron temperatures (see next section).
Despite the small number, the three \othreea\ detections are found in 
galaxies where $W_e(\hbeta) \ga 40$~\AA\ and 
$I(\hbeta) \ga 7 \times 10^{-15}$ erg~s$^{-1}$~cm$^{-2}$.
\begin{table}
\scriptsize 
\begin{center}
\renewcommand{\arraystretch}{1.}
\caption{
Derived properties.
Col.~(1): \ion{H}{II} region.
Col.~(2): \hbeta\ intensity, corrected for underlying Balmer
absorption and adopted reddening. 
Cols.~(3) and (4): Computed and adopted reddening values.
Col.~(5): Observed \hbeta\ emission equivalent width.
Col.~(6): Electron density.
Col.~(7): O$^{+2}$ electron temperature arising from
\othreea\ detection or upper limit.
Col.~(8): O$^+$ electron temperature.
NOTE:
$^a$~Background galaxy with $v_{\odot}/c \simeq +0.11$.
Observed (uncorrected) \hbeta\ flux and \hbeta\ emission equivalent
width are listed.
%
\label{table_derivedprops}
}
\begin{tabular}{lccccccc}
\hline \hline
& & Derived & Adopted & & & & \\
& $I($H$\beta)$ & $E(B-V)$ & $E(B-V)$ & 
$W_{\rm em}$(\hbeta) & $n_e$ & $T_e$(O$^{+2}$) & $T_e$(O$^+$) \\
H II Region & (ergs s$^{-1}$ cm$^{-2}$) & (mag) & (mag) & 
(\AA) & (cm$^{-3}$) & (K) & (K) \\
(1) & (2) & (3) & (4) & (5) & (6) & (7) & (8) \\
\hline
\multicolumn{8}{c}{{\sf Centaurus A group dwarfs}} \\ 
\hline
A1243$-$335 \ion{H}{II}\#A & $(4.44 \pm 0.54) \times 10^{-16}$ &
        $+0.14 \pm 0.16$ & $+0.14$ & $18.3 \pm 2.1$ & 100 &
        \ldots & \ldots \\
A1324$-$412 ap1 & $(1.88 \pm 0.34) \times 10^{-15}$ &
        $+0.14 \pm 0.14$ & $+0.14$ & $14.7 \pm 2.4$ & 100 &
        \ldots & \ldots \\
A1334$-$277 ap1 & $(1.15 \pm 0.19) \times 10^{-16}$ &
        $-0.07 \pm 0.25$ & 0 & $17.2 \pm 2.5$ & 100 &
        \ldots & \ldots \\
A1346$-$358 \ion{H}{II}\#A & $(7.73 \pm 0.34) \times 10^{-15}$ & 
        $+0.427 \pm 0.087$ & $+0.427$ & $103.7 \pm 4.3$ & 100 &
        10935 & 11504 \\
DDO 161 ap1 & $(3.71 \pm 0.18) \times 10^{-15}$ &
        $+0.399 \pm 0.091$ & $+0.399$ & $43.6 \pm 1.4$ & 31 &
        $<$ 15460 & $<$ 13970 \\
NGC 5264 ap1 & $(5.66 \pm 0.52) \times 10^{-16}$ &
        $+0.15 \pm 0.14$ & $+0.15$ & $11.09 \pm 0.84$ & 100 &
        \ldots & \ldots \\
\hline
\multicolumn{8}{c}{{\sf Sculptor group dwarfs}} \\ 
\hline
AM0106$-$382 ap1 & $(3.30 \pm 0.40) \times 10^{-16}$ &
        \ldots & 0 & $24.3 \pm 2.7$ & 100 & \ldots & \ldots \\
AM0106$-$382 ap2 & $(9.28 \pm 0.82) \times 10^{-16}$ & 
        \ldots & 0 & $18.4 \pm 1.3$ & 100 & \ldots & \ldots \\
AM0106$-$382 ap3 & $(5.86 \pm 0.52) \times 10^{-16}$ & 
        \ldots & 0 & $15.4 \pm 1.0$ & 100 & \ldots & \ldots \\
ESO347$-$G017 ap1 & $(2.46 \pm 0.14) \times 10^{-15}$ &
        $+0.02 \pm 0.11$ & $+0.02$ & $17.91 \pm 0.45$ & 100 &
        $<$ 13680 & $<$ 13090 \\
ESO347$-$G017 ap2 & $(2.33 \pm 0.15) \times 10^{-15}$ &
        $+0.14 \pm 0.12$ & $+0.14$ & $14.86 \pm 0.54$ & 100 &
        \ldots & \ldots \\
ESO347$-$G017 ap3 & $(8.1 \pm 1.0) \times 10^{-16}$ &
        $+0.25 \pm 0.19$ & $+0.25$ & $12.5 \pm 1.3$ & 100 &
        \ldots & \ldots \\
ESO348$-$G009 ap1 & $(9.8 \pm 1.4) \times 10^{-16}$ &
        \ldots & 0 & $16.6 \pm 2.2$ & 100 & \ldots & \ldots \\
ESO348$-$G009 ap2$\,^a$ & 
        $(2.32 \pm 0.76) \times 10^{-15}$ &
        \ldots & \ldots & $5.2 \pm 1.7$ & \ldots & \ldots &
        \ldots \\
UGCA 442 \ion{H}{II}\#2 & $(2.23 \pm 0.13) \times 10^{-15}$ &
        $+0.13 \pm 0.11$ & $+0.13$ & $17.08 \pm 0.47$ & 100 &
        $<$ 17480 & $<$ 14830 \\
\hline
\multicolumn{8}{c}{{\sf Other southern dwarfs}} \\ 
\hline
A0355$-$465 \ion{H}{II}\#B & $(2.70 \pm 0.14) \times 10^{-15}$ &
        $+0.318 \pm 0.095$ & $+0.318$ & $50.2 \pm 2.3$ & 100 &
        \ldots & \ldots \\
ESO358$-$G060 ap1 & $(1.35 \pm 0.11) \times 10^{-15}$ &
        $-0.13 \pm 0.15$ & 0 & $56.8 \pm 4.3$ & 100 &
        $<$ 18160 & $<$ 15090 \\
IC 1613 \ion{H}{II}\#13 & $(4.25 \pm 0.18) \times 10^{-15}$ &
        \ldots & 0 & $89.1 \pm 4.6$ & 100 & $<$ 14540 & $<$ 16760 \\
IC 1613 \ion{H}{II}\#37 & $(1.275 \pm 0.068) \times 10^{-14}$ &
        $+0.10 \pm 0.11$ & $+0.10$ & $514 \pm 97$ & 100 &
        17910 & 14990 \\
IC 2032 ap1 & $(1.94 \pm 0.22) \times 10^{-16}$ &
        $+0.01 \pm 0.16$ & $+0.01$ & $15.0 \pm 1.5$ & 100 &
        \ldots & \ldots \\ 
IC 5152 \ion{H}{II}\#A & $(1.732 \pm 0.091) \times 10^{-14}$ &
        \ldots & 0 & $44.90 \pm 0.85$ & 37 & 
        12360 & 12360 \\ 
NGC 2915 ap1 & $(1.84 \pm 0.21) \times 10^{-14}$ &
        $+0.72 \pm 0.16$ & $+0.72$ & $4.94 \pm 0.26$ & 100 &
        \ldots & \ldots \\
NGC 2915 ap2 & $(7.62 \pm 0.88) \times 10^{-16}$ &
        $+0.36 \pm 0.17$ & $+0.36$ & $13.8 \pm 1.4$ & 100 &
        \ldots & \ldots \\
NGC 3109 \ion{H}{II}\#6 ap1 & $(2.67 \pm 0.59) \times 10^{-16}$ &
        $-0.19 \pm 0.33$ & 0 & $11.9 \pm 2.3$ & 100 &
        \ldots & \ldots \\
NGC 3109 \ion{H}{II}\#6 ap2 & $(2.99 \pm 0.65) \times 10^{-16}$ &
        $-0.15 \pm 0.31$ & 0 & $7.3 \pm 1.2$ & 100 & 
        \ldots & \ldots \\
NGC 3109 \ion{H}{II}\#6 ap3 & $(9.91 \pm 0.83) \times 10^{-16}$ & 
        $+0.03 \pm 0.16$ & $+0.03$ & $21.0 \pm 1.2$ & 100 & 
        \ldots & \ldots \\
NGC 3109 \ion{H}{II}\#6 ap4 & $(2.46 \pm 0.18) \times 10^{-15}$ &
        $+0.06 \pm 0.14$ & $+0.06$ & $23.25 \pm 0.80$ & 100 &
        \ldots & \ldots \\
NGC 3109 \ion{H}{II}\#6 ap5 & $(4.58 \pm 0.33) \times 10^{-15}$ &
        $+0.08 \pm 0.14$ & $+0.08$ & $31.3 \pm 1.1$ & 100 &
        \ldots & \ldots \\
NGC 3109 \ion{H}{II}\#6 ap6 & $(1.92 \pm 0.14) \times 10^{-15}$ & 
        $+0.08 \pm 0.14$ & $+0.08$ & $67.5 \pm 4.5$ & 100 &
        \ldots & \ldots \\
NGC 3109 \ion{H}{II}\#6 ap7 & $(6.2 \pm 1.3) \times 10^{-16}$ &
        $+0.28 \pm 0.30$ & $+0.28$ & $20.7 \pm 4.3$ & 100 &
        \ldots & \ldots \\
Sag DIG \ion{H}{II}\#3 & $(8.82 \pm 0.57) \times 10^{-16}$ & 
        $+0.23 \pm 0.12$ & $+0.23$ & $67.0 \pm 4.8$ & 100 &
        \ldots & \ldots \\
\hline
\end{tabular}
\end{center}
\end{table}

\section{Nebular Abundances}
\label{sec_abundderive}

\subsection{Oxygen Abundances: Direct Method}
\label{sec_standard}

The direct or standard method of obtaining oxygen abundances 
from emission lines is applicable to any galaxy where \othreea\ is
detectable and for which the doubly ionized O$^{+2}$ ion is the
dominant form of oxygen \citep{osterbrock}. 
A summary of the ``standard'' method by which oxygen abundances are
derived can be found in \cite{dinerstein90}.
Computations were performed with SNAP.
The relative abundances of singly- and doubly-ionized oxygen and the
total oxygen abundance by number are computed using the 
method described by \cite{lee03}. 
An O$^{+2}$/H abundance was computed using an O$^{+2}$ temperature,
derived from the intensity of the \othreea\ and \othree\ lines, 
and an O$^+$/H abundance was computed using an O$^+$ temperature
derived using Eq.~(2) from \cite{itl97}.

Direct (\othreea) abundances were obtained for three galaxies
(A1346$-$358, IC~1613 \ion{H}{II}\#37, and IC~5152 \ion{H}{II}\#A)
and are listed in Table~\ref{table_nebabund}.
Errors in direct oxygen abundances were computed from the maximum and
minimum possible values, given the errors in the line intensities;
errors in reddening and temperature are not included.
For the remaining galaxies, secondary techniques using the
bright emission lines of ionized oxygen are utilized to derive
oxygen abundances.
\begin{table}
\scriptsize 
\begin{center}
\renewcommand{\arraystretch}{1.}
\caption{
Derived nebular abundances.
Col.~(1): \ion{H}{II} region.
Col.~(2): Oxygen abundance from \othreea\ measurements; lower limits 
to the oxygen abundance were obtained from 2$\sigma$ upper limits
to the \othreea\ flux.
Col.~(3): Oxygen abundance derived using bright-line method
by \cite{mcgaugh91}.
Col.~(4): Oxygen abundance derived using bright-line method 
by \cite{pilyugin00}.
Cols.~(5) and (6): Nitrogen-to-oxygen abundance ratios:
derived with O$^+$ temperatures from Table~\ref{table_derivedprops} and
with the bright-line method, respectively.
Cols.~(7) and (8): Neon-to-oxygen abundance ratios:
derived with O$^{+2}$ temperatures from Table~\ref{table_derivedprops} and
with the bright-line method, respectively.
NOTES:
$^a$~The Pilyugin value is uncertain, but upper branch abundances
appear to be correct; see text in Sect.~\ref{sec_ngc5264}.
$^b$~Computed using the method of \cite{thurston96}.
$^c$~$I(\ntwo)$ is likely a lower limit.
$^d$~$I(\nethree)$ likely too low.
$^e$~Pilyugin value is uncertain. 
Oxygen abundances derived with $I(\ntwob)/I(\halpha)$
calibrations \citep{vanzee98sp,denicolo02} are consistent with lower
branch values; the value listed here is the average.
$^f$~\ntwob\ not measured; cannot break degeneracy in bright-line
method. 
$^g$~Same as note~$e$, but see also Sect.~\ref{sec_ngc3109}.
%
\label{table_nebabund}
}
\begin{tabular}{lccccccc}
\hline \hline
& \multicolumn{3}{c}{12$+$log(O/H)} & & & \\
\cline{2-4}
& Direct & \multicolumn{2}{c}{Bright-Line} & 
\multicolumn{2}{c}{log(N/O)} &
\multicolumn{2}{c}{log(Ne/O)} \\ 
\cline{5-8}
\ion{H}{II} Region & [O III]$\lambda$4363 & McGaugh & Pilyugin & 
Direct & Bright-line & Direct & Bright-line \\
(1) & (2) & (3) & (4) & (5) & (6) & (7) & (8) \\
\hline
\multicolumn{8}{c}{{\sf Centaurus A group dwarfs}} \\
\hline
A1243$-$335 \ion{H}{II}\#A & \ldots & 7.87 & 7.69 & \ldots & $-1.62$ &
\ldots & \ldots\\  
A1334$-$277 ap1 & \ldots & 7.45 & 7.34 & \ldots & $-1.04$ & \ldots &
\ldots \\ 
A1346$-$358 \ion{H}{II}\#A & $8.19 \pm 0.06$ & 8.26 & 8.22 & $-1.37
\pm 0.08$ & $-1.44$ & $-0.46 \pm 0.05$ & $-0.50$ \\ 
DDO 161 ap1 & $>$ 7.76 & 8.08 & 8.03 & \ldots & $-1.91$ & \ldots &
\ldots \\ 
NGC 5264 ap1 & \ldots & 8.66 & 8.54$\,^a$ & \ldots & $-0.57\,^b$
& \ldots & \ldots \\ 
\hline
\multicolumn{8}{c}{{\sf Sculptor group dwarfs}} \\
\hline
AM0106$-$382 ap1 & \ldots & 7.54 & 7.71 & \ldots & $-1.22$ & \ldots &
\ldots \\ 
AM0106$-$382 ap2 & \ldots & 7.58 & 7.56 & \ldots & $-1.47$ & \ldots &
\ldots \\ 
AM0106$-$382 ap3 & \ldots & 7.61 & 7.59 & \ldots & $-1.58$ & \ldots &
\ldots \\ 
ESO347$-$G017 ap1 & $>$ 7.80 & 7.89 & 7.78 & \ldots & $-1.30$ & \ldots
& $-0.52$\\ 
ESO347$-$G017 ap2 & \ldots & 7.92 & 7.76 & \ldots & $-1.31$ & \ldots &
$-0.53$ \\ 
ESO347$-$G017 ap3 & \ldots & 7.96 & 8.03 & \ldots & $> -1.86\,^c$ &
\ldots & \ldots \\ 
ESO348$-$G009 ap1 & \ldots & 7.89 & 8.07 & \ldots & $-1.60$ & \ldots &
\ldots \\ 
UGCA 442 \ion{H}{II}\#2 & $>$ 7.48 & 7.81 & 7.88 & \ldots & $-1.41$ &
\ldots & \ldots \\ 
\hline
\multicolumn{8}{c}{{\sf Other southern dwarfs}} \\
\hline
A0355$-$465 \ion{H}{II}\#B & \ldots & 8.23 & 8.01 & \ldots & $-1.61$ &
\ldots & $-0.48$ \\ 
ESO358$-$G060 ap1 & $>$ 7.26 & 7.38 & 7.26 & \ldots & $-1.24$ & \ldots
& $> -1.11\,^d$ \\ 
IC 1613 \ion{H}{II}\#13 & $>$ 7.61 & 7.90 & 7.89 & \ldots & $-1.40$ &
\ldots & \ldots \\ 
IC 1613 \ion{H}{II}\#37 & $7.62 \pm 0.05$ & 7.88 & 7.71 & $-1.13 \pm
0.18$ & $-1.35$ & $-0.60 \pm 0.05$ & $-0.62$ \\ 
IC 2032 ap1 & \ldots & 7.96 & 7.98 & \ldots & $-1.37$ & \ldots &
$-0.21$ \\ 
IC 5152 \ion{H}{II}\#A & $7.92 \pm 0.07$ & 7.91 & 7.80 & $-1.05 \pm
0.12$ & $-1.09$ & $-0.69 \pm 0.08$ & $-0.77$ \\ 
NGC 2915 ap1 & \ldots & 8.29 & 8.33$\,^e$ & \ldots & $-1.71$ & \ldots &
\ldots \\ 
NGC 2915 ap2 & \ldots & 8.21 & 8.35 & \ldots & $-1.25$ & \ldots &
\ldots \\ 
NGC 3109 \ion{H}{II}\#6 ap1$\,^f$ & \ldots & \ldots & \ldots & \ldots &
\ldots & \ldots & \ldots \\ 
NGC 3109 \ion{H}{II}\#6 ap2$\,^f$ & \ldots & \ldots & \ldots & \ldots &
\ldots & \ldots & \ldots \\ 
NGC 3109 \ion{H}{II}\#6 ap3$\,^g$ & \ldots & 7.50 & 7.40 &
\ldots & $-$1.37 & \ldots & \ldots \\
NGC 3109 \ion{H}{II}\#6 ap4$\,^g$ & \ldots & 8.07 & 8.13 &
\ldots & $-$1.36 & \ldots & \ldots \\ 
NGC 3109 \ion{H}{II}\#6 ap5$\,^g$ & \ldots & 7.64 & 7.52 &
\ldots & $-$1.28 & \ldots & \ldots \\ 
NGC 3109 \ion{H}{II}\#6 ap6$\,^g$ & \ldots & 7.60 & 7.51 &
\ldots & $-$1.20 & \ldots & \ldots \\ 
NGC 3109 \ion{H}{II}\#6 ap7$\,^g$ & \ldots & 7.85 & 8.08 &
\ldots & $-$1.38 & \ldots & \ldots \\ 
Sag DIG \ion{H}{II}\#3 & \ldots & 7.44 & 7.33 & \ldots & $-1.63$ &
\ldots & \ldots \\ 
\hline
\end{tabular}
\end{center}
\end{table}

\subsection{Oxygen Abundances: Bright-Line Method}
\label{sec_brightline}

In the absence of \othreea, the bright-line or empirical method
was used to compute oxygen abundances.
The method is so called because the oxygen abundance is given in terms
of the bright [O~II] and [O~III] lines. 
\cite{pagel79} suggested that the ratio
\begin{equation}
R_{23} = \frac{I({\otwo}) + I({\othree})}{I({\hbeta})}
\label{eqn_r23_def}
\end{equation}
could be used as an abundance indicator.
For the dwarf galaxies in the present work, the calibrations
by \cite{mcgaugh91,mcgaugh94} and \cite{pilyugin00} were used to
derive oxygen abundances.

\begin{figure*}
\centering
\includegraphics[width=9.cm]{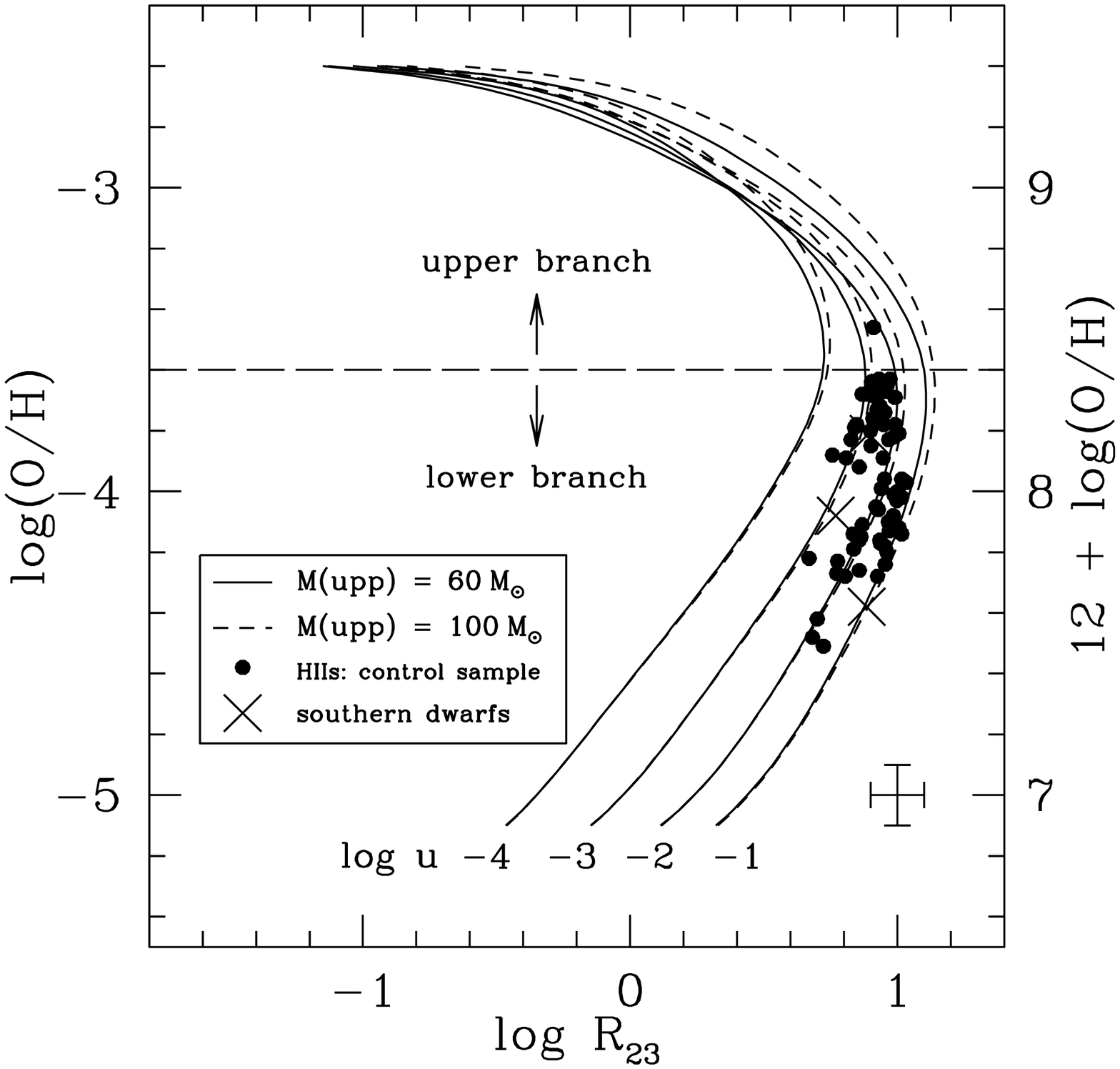} 
\caption{
Oxygen abundance versus bright-line indicator, $R_{23}$.
The filled circles indicate \ion{H}{II} regions from the
control sample of nearby dIs \citep{lee03}, whose oxygen abundances
were obtained directly from measurements of the \othreea\ emission
line.
Crosses indicate southern dwarfs in the present sample for which
\othreea\ was measured.
Calibration curves for log(O/H) against log~$R_{23}$
(McGaugh 1997, private communication) are plotted
for four different values of the ionization parameter 
(log $u$; see \citealp{mcgaugh91}).
These curves were derived using a standard stellar initial mass
function for a cluster of ionizing stars with an upper mass limit of 
${\cal M}_{\rm upp}$ = 60~\msun\ (solid lines) and 
100~\msun\ (short-dash lines); see also \cite{mcgaugh91}.
The horizontal long-dash line marks the approximate boundary below
(above) which the lower (upper) branch occurs. 
\othreea\ abundances are consistent with lower branch values.
The error bars at the lower right indicates typical uncertainties of
0.1~dex for the $R_{23}$ indicator and 0.1~dex for the direct
oxygen abundance.
}
\label{fig_oxyr23}
\end{figure*}

\cite{mcgaugh91,mcgaugh94} produced a set of photoionization models
using $R_{23}$ and 
\begin{equation}
O_{32} = \frac{I({\othree})}{I({\otwo})}
\end{equation}
to estimate the oxygen abundance. 
However, $R_{23}$ is not a monotonic function of the oxygen abundance.
At a given value of $R_{23}$, there are two possible choices of the
oxygen abundance as shown in Fig.~\ref{fig_oxyr23}.
Each filled circle represents an \ion{H}{II} region from dIs in the 
control sample with measured \othreea.
Model curves from \cite{mcgaugh91} are superimposed.
A long-dashed line marks the approximate boundary below (above) which 
the lower branch (upper branch) occurs.
Despite the fact the oxygen abundances for \ion{H}{II} regions in
the control sample of dIs range from about one-tenth to about one-half
of the solar value, these \ion{H}{II} regions are clustered
around the ``knee'' of the curves, where ambiguity is greatest
about the choice of the appropriate branch in the absence of
\othreea.

\begin{figure*}
\centering
\includegraphics[width=9.cm]{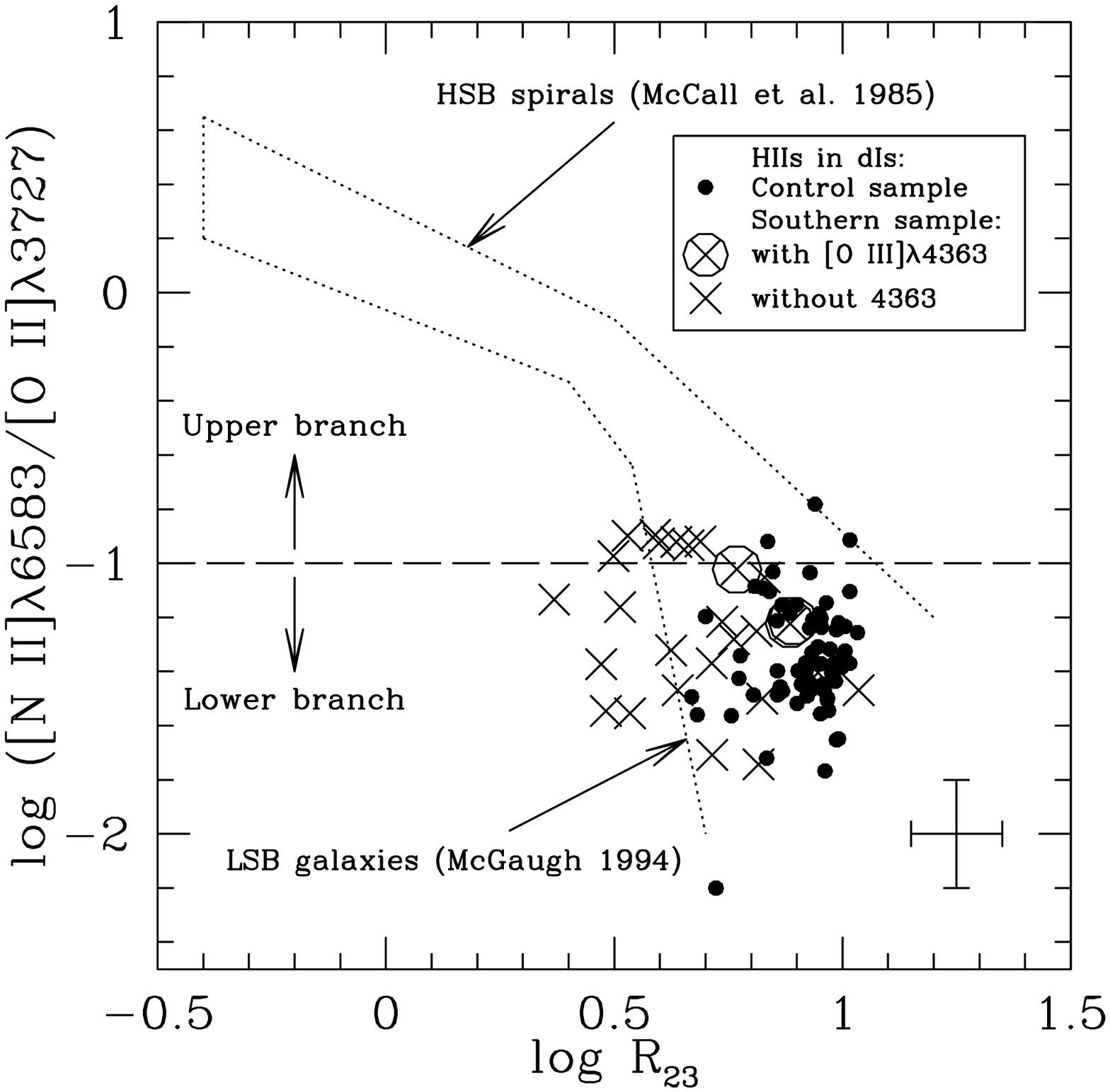} 
\caption{
\ntwootwo\ discriminant versus bright-line indicator, $R_{23}$.
Filled circles indicate \ion{H}{II} regions from the control sample of 
nearby dIs with \othreea\ detections \citep{lee03}.
Crosses indicate \ion{H}{II} regions for the present sample
of southern dwarfs; crosses surrounded by open circles indicate
\othreea\ measurements.
The error bars at the lower right indicate typical uncertainties 
of 0.1~dex for the $R_{23}$ indicator and 0.2~dex (at most) for the
\ntwootwo\ ratio.
The dotted lines mark the regions in the diagram occupied by
high surface brightness (HSB) spiral galaxies \citep{mrs85} at the
upper left and low surface brightness (LSB) galaxies \citep{mcgaugh94}
towards the lower right.
As suggested by McGaugh, the horizontal dashed line marks the 
approximate boundary below (above) which the lower (upper) branch of
the bright-line method is selected to determine a unique value of
an oxygen abundance. 
}
\label{fig_n2o2r23}
\end{figure*}

Fortunately, $I$(\ntwob)/$I$(\otwo), or the \ntwootwo\ intensity ratio
can discriminate between the lower and upper branches
\citep{mrs85,mcgaugh91,mcgaugh94}.
The strength of the \ntwob\ line is roughly proportional to the
nitrogen abundance and the \ntwootwo\ intensity ratio is relatively
insensitive to ionization.
\cite{mcgaugh94} has shown that in galaxies ranging from sub-solar
to solar metallicities, \ntwootwo\ can vary by one to two orders of
magnitude and that \ntwootwo\ is roughly below (above) 0.1 at 
low (high) oxygen abundance.
A plot of the \ntwootwo\ intensity ratio versus $R_{23}$ is shown in
Fig.~\ref{fig_n2o2r23}.
While most \ion{H}{II} regions in the present sample lie in the lower
branch regime, they lie to the left of the locus of points defined by
\ion{H}{II} regions in nearby dIs.  
This is expected for metal-poor galaxies as ionization effects
become more prominent \citep{mcgaugh91}.
The corrected \ntwootwo\ was used to determine the branch for
computing the oxygen abundance in both the McGaugh and Pilyugin
calibrations.

For the McGaugh (1997, private communication) calibration,
analytical equations for the oxygen abundance are given in
terms of $x \equiv \log\,R_{23}$ and $y \equiv \log\,O_{32}$.
The expressions for lower branch and upper branch oxygen abundances
are
\begin{eqnarray}
12 + \log\,({\rm O/H})_{\rm lower} & = & 
12 - 4.93 + 4.25x - 3.35\sin \,(x) - 0.26y - 0.12\sin \,(y), \\
12 + \log\,({\rm O/H})_{\rm upper} & = & 
12 - 2.65 - 0.91x + 0.12y \, \sin \,(x),
\end{eqnarray}
respectively, where the argument of the trigonometric 
function is in radians. 

Pilyugin suggested a new calibration for the bright-line
method. 
His method at low metallicities accounts for the systematic
uncertainties in the $R_{23}$ method, whereas at high metallicities,
he obtains a relation for the oxygen abundance as a function of the
intensities of the bright [O~II] and [O~III] lines.
For lower branch and upper branch abundances, we use Eq.~(4) from 
\cite{pilyugin00}, and Eq.~(8) from \cite{pilyugin01a},
respectively.

\begin{figure*}
\centering
\includegraphics[width=9.cm]{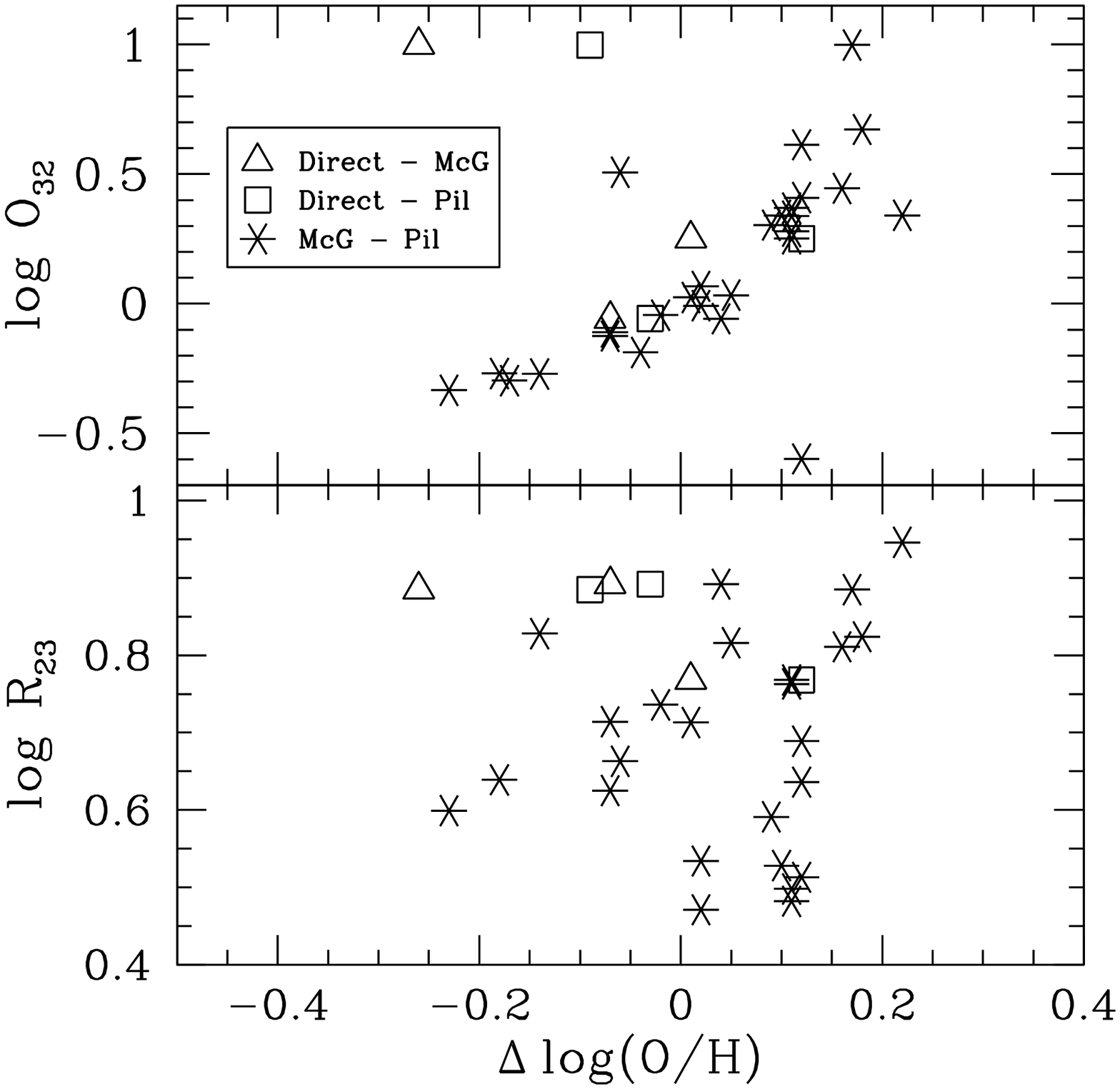} 
\caption{
Difference in oxygen abundance from various methods
versus log~$O_{32}$ (top panel), and 
versus log~$R_{23}$ (bottom panel).
Each point represents an \ion{H}{II} region from dwarf galaxies
in the southern sample.
``Direct'' denotes oxygen abundances derived from \othreea\
measurements, ``McG'' denotes oxygen abundances derived using
the bright-line method by \cite{mcgaugh91}, and ``Pil'' denotes
oxygen abundances derived using the bright-line method 
by Pilyugin \citeyearpar{pilyugin00,pilyugin01a}.
}
\label{fig_oxydiff}
\end{figure*}

Bright-line oxygen abundances are derived and listed in
Table~\ref{table_nebabund}.
Fig.~\ref{fig_oxydiff} shows how the different determinations
of the oxygen abundance vary with $O_{32}$ and $R_{23}$.
Differences in derived oxygen abundance between the direct (\othreea)
and bright-line McGaugh methods, between the direct and bright-line 
Pilyugin methods, and between the two bright-line methods are shown.
The separations among the three methods appear to increase with 
increasing $O_{32}$.
The difference between the McGaugh and Pilyugin calibrations 
(indicated by asterisks) appears to correlate with log $O_{32}$; 
this effect is also observed by \cite{scm03}. 
IC~1613 \ion{H}{II}\#37 with the largest measured $O_{32}$
($\log\,O_{32} = 0.998$) exhibits the largest overall discrepancy
among the direct, McGaugh, and Pilyugin methods (12$+$log(O/H) = 7.62,
7.88, and 7.71, respectively). 

\subsection{Nitrogen-to-Oxygen and Neon-to-Oxygen Abundance Ratios}
\label{sec_noratio}

Based upon observations of \ion{H}{II} regions in spiral and dwarf
galaxies, nitrogen appears to be both a primary and secondary product
of nucleosynthesis. 
It remains uncertain, however, whether nitrogen is produced mostly
from short--lived massive stars or from longer--lived
intermediate--mass stars.
An extensive review of the possible origins for nitrogen
is discussed by \cite{henry00}.

Measurements of the nitrogen-to-oxygen ratio, N/O, have been used to
differentiate between the different origins for nitrogen.
It has been suggested that N/O can be used as a ``clock'' to
measure the time since the last burst of star formation 
(e.g., \citealp{garnett90,sbk97,scm03}).
This scenario works if bursts of star formation are separated by long 
quiescent periods, if the delivery of nitrogen into the interstellar
gas is delayed relative to oxygen, and if there is no significant
metals loss. 
The result is that N/O values are low at a given O/H if a burst of
star formation has occurred recently, whereas N/O values are high
after a long quiescent period.

For low-abundance \ion{H}{II} regions, 
$\log({\rm N}/{\rm O}) \approx \log({\rm N}^+/{\rm O}^+)$
is a good approximation (e.g., \citealp{garnett90}).
Assuming that $T_e({\rm N}^+) = T_e({\rm O}^+)$,
the nitrogen-to-oxygen abundance ratio is
\begin{equation}
{\rm N}/{\rm O} =
\frac{I({\rm [N\;II]}\lambda\,6583)}{I({\rm [O\;II]}\lambda\,3727)} \cdot 
\frac{j({\rm [O\;II]}\lambda\,3727;\;n_e,\,T_e({\rm O}^+))}
{j({\rm [N\;II]}\lambda\,6583;\;n_e,\,T_e({\rm O}^+))}, \\
\label{eqn_no_direct}
\end{equation}
In the absence of \othreea, a bright-line oxygen abundance
was derived using the McGaugh calibration and an O$^{+2}$
temperature was obtained using the correlation between oxygen
abundance and temperature in Fig.~7 of \cite{mcgaugh91}.
Subsequently, an O$^+$ temperature was estimated using
the data in Table~1 of \cite{vce93}.
The nitrogen-to-oxygen abundance ratio was then computed by using
the appropriate line intensities with Eq.~(9) from
\cite{pagel92}. 
%

Neon is a product of $\alpha$-processes in nucleosynthesis
occurring in the same massive stars which produce oxygen.
As a result, the neon-to-oxygen ratio, Ne/O, is expected to be
constant with oxygen abundance. 
Assuming that doubly-ionized neon is found in the same zone
as doubly-ionized oxygen and that 
$T_e({\rm Ne}^{+2}) = T_e({\rm O}^{+2})$ and that 
$\log({\rm Ne}/{\rm O}) \simeq \log({\rm Ne}^{+2}/{\rm O}^{+2})$,
the neon-to-oxygen abundance ratio is
\begin{equation}
{\rm Ne}/{\rm O} =
\frac{I({\rm [Ne\;III]}\lambda\,3869)}{I({\rm [O\;III]}\lambda\,5007)} \cdot 
\frac{j({\rm [O\;III]}\lambda\,5007;\;n_e,\,T_e({\rm O}^{+2}))}
{j({\rm [Ne\;III]}\lambda\,3869;\;n_e,\,T_e({\rm Ne}^{+2}))}.
\label{eqn_neo_direct}
\end{equation}
In the absence of \othreea, a bright-line temperature for the O$^{+2}$
zone is obtained (see previous paragraph) and the neon-to-oxygen
abundance ratio is computed using the appropriate line intensities
with Eq.~(10) from \cite{pagel92}.
%

Table~\ref{table_nebabund} lists N/O and Ne/O abundance ratios.
For the three galaxies with \othreea\ detections, N/O and Ne/O values
derived using the direct method generally agree with those derived
with the bright-line method, except that the direct value of log(N/O)
for IC~1613 \ion{H}{II}\#37 is 0.2~dex larger than the 
bright-line value.
A few of the Ne/O values are higher than those derived for blue
compact dwarf galaxies: log(Ne/O) $\approx$ $-$0.7 \citep{it99}, 
but \nethree\ flux measurements for galaxies in the present sample
may be overestimated from noisy spectra.

\section{Discussion}
\label{sec_discuss}

For bright-line determinations of oxygen abundances, the reported 
values for each galaxy are means of values derived from the McGaugh
and the Pilyugin calibrations. 
New oxygen abundances are reported for eight galaxies from the 
southern sample: 
in the Cen~A group, A1334$-$277, DDO~161, and NGC~5264; 
in the Scl group, AM0106$-$382, ESO347$-$G017
(but see Appendix~\ref{sec_app}), and ESO348$-$G009; 
and finally, ESO358$-$G060, IC~2032, and NGC~2915.

\subsection{Comments on Individual Galaxies}
\label{sec_individ}

All of the galaxies in the southern sample are presented
below in alphabetical order.

\subsubsection{A0355$-$465 (ESO249$-$G032)}

\cite{webster83} obtained spectra of two \ion{H}{II} regions 
(\#A and \#B).
Unfortunately, their \othreea\ measurement in \ion{H}{II}
region \#B is very uncertain with their quoted error in excess
of 40\%; they derived 12$+$log(O/H) = 8.41 and log(N/O) = $-$1.70.
From our spectrum of \ion{H}{II} region \#B, we obtain 12$+$log(O/H) =
8.12 and log(N/O) = $-$1.61 using the bright-line method.
While our $I$(\othree)/$I$(\hbeta) ratio is about 80\% of that
reported by \cite{webster83}, our $I$(\otwo)/$I$(\hbeta) ratio is
about a factor of two lower.

There is currently no velocity listed for this galaxy in the NED
database. 
From the emission lines in the spectrum, the heliocentric
velocity was estimated to be 1168 km~s$^{-1}$.
Assuming that A0355$-$465 is isolated and that the Hubble constant 
has a value of 75 km~s$^{-1}$~Mpc$^{-1}$, the estimated distance is
15.6~Mpc and the distance modulus is 31.0.
This distance is comparable to that of the Virgo Cluster and the
Fornax Cluster. 
The estimated absolute magnitude in $B$ is $-$14.7, which 
is similar to that of the Local Group dI NGC~6822.
The angular dimensions from the NED database are 2\farcm3 by 1\farcm4,
corresponding to linear dimensions 10.4~kpc by 6.3~kpc.
This is comparable to the dimensions of the Magellanic dwarf NGC~4532
(VCC~1554) in the Virgo Cluster \citep{hoffman99}.

An upper limit to the total 21-cm flux was obtained by 
\cite{longmore82}, though another measurement should be obtained
for confirmation.
Nevertheless, A0355$-$465 exhibits the lowest $M_{\rm HI}/L_B$ of the
present sample.
This dwarf lies 9\farcm6 away from the face-on SB(rs)cd galaxy
NGC~1493, which has a measured heliocentric velocity of 
1054 km~s$^{-1}$, roughly similar to our estimated velocity for 
A0355$-$465.
In fact, the dwarf may be a member of a very small group of galaxies
with at least four catalogued members \citep[LGG~106;][]{garcia93}.

\subsubsection{A1243$-$335 (ESO381$-$G020)}

\cite{webster83} identified three \ion{H}{II} regions in this Cen~A
group dwarf irregular galaxy.
The brightest (labelled \#A) was found in the southeast corner of the
galaxy; \ion{H}{II} regions \#B and \#C were found in the northwest
and southwest part of the galaxy, respectively.
They obtained spectroscopy for all three \ion{H}{II} regions and
detected \othreea\ only in \ion{H}{II} \#A.
While their quoted uncertainty for the \othreea\ detection
is rather large ($>$ 40\%), their derived oxygen abundance was
12$+$log(O/H) = 8.04.
\ion{H}{II} regions \#A and \#B appear to be spatially coincident 
with local maxima in \ion{H}{I} \citep{cote00}.
The \ion{H}{I} extent is 4.5 times larger than the optical galaxy.

Our spectrum of \ion{H}{II} region \#A revealed the bright
oxygen lines, but no \othreea.
The derived bright-line oxygen abundance is 12$+$log(O/H) = 7.78,
which is almost 0.3~dex lower than the \cite{webster83} value.
The adopted nitrogen-to-oxygen abundance ratio is
log(N/O) = $-1.62$.

\subsubsection{A1324$-$412 (ESO324$-$G024)}

For this Cen~A group dwarf irregular, \cite{cote97} measured the total
\ion{H}{I} flux, and \cite{kara02b} derived a TRGB distance.
However, our spectrum only revealed \hbeta\ and \halpha\
emission; the subsequent derived reddening is $E(B-V) = +0.14$ mag.

\subsubsection{A1334$-$277 (ESO444$-$G084)}

The oxygen abundance reported here (12$+$log(O/H) = 7.40)
is the first ever published for this Cen~A group member
and is the third most metal-poor galaxy in the present study.
While the upper limit to the \ntwootwo\ intensity ratio may indicate
an upper branch abundance for this Cen~A group dwarf, a low
metallicity is assumed, because the $I$(\ntwob)/$I$(\halpha) and 
$I$(\stwoa)/$I$(\halpha) ratios are small, i.e., less than 3~\% and
2~\%, respectively.  
Indeed, bright-line calibrations using $I$(\ntwob)/$I$(\halpha)
\citep{vanzee98sp,denicolo02} yield lower branch abundances.

While this dwarf galaxy is relatively metal-poor,
it is not particularly isolated.
The TRGB distance of this dI is 4.61~Mpc, which puts this galaxy
380~kpc from the spiral galaxy M~83 \citep{kara02b}.
While the \ion{H}{I} extent is almost six times larger than the
optical extent \citep{cote00}, the $M_{\rm HI}/L_B$ ratio is normal
for its luminosity compared to typical dIs
(e.g., \citealp{rh94,skillman96,pildis97}). 

\subsubsection{A1346$-$358 (ESO383$-$G087)}

\cite{webster83} obtained spectroscopy of three \ion{H}{II} regions
(A, B, and C), but did not detect \othreea\ at all.
They estimated the oxygen abundance to be 
12$+$log(O/H) $\approx$ 8.3.
In our spectrum of \ion{H}{II} region \#A, \othreea\ was detected and
the subsequent oxygen abundance is 12$+$log(O/H) = $8.19 \pm 0.06$, or
about one-fifth of the solar value. 
The nitrogen- and neon-to-oxygen ratios are:
log(N/O) = $-1.37 \pm 0.08$ and log(Ne/O) = $-0.46 \pm 0.05$.
The latter is about 0.2~dex larger than values obtained
for blue compact dwarf galaxies at similar oxygen abundance
\citep{it99}.
This dwarf lies (in projection) approximately between the elliptical
Cen~A and the spiral M~83.
The proximity to these two giant galaxies may explain why the
$M_{\rm HI}/L_B$ value for the dwarf is several times lower
than typical values for dIs at comparable luminosities
(e.g., \citealp{sdk92}).

\subsubsection{AM0106$-$382}

Aside from the properties listed in Table~\ref{table_gxylist}, and
the inclusion of this galaxy in the \ion{H}{I} catalog by
\cite{cote97}, there is little in the literature describing further
this dwarf irregular in the Sculptor group. 
The adopted oxygen abundance is 12$+$log(O/H) = 7.60
and the nitrogen-to-oxygen ratio is log(N/O) = $-1.42$.
These values are the first ever published for this galaxy.

\subsubsection{DDO 161}

This Cen~A group dwarf galaxy was studied in some detail by
\cite{kkm81}, but the oxygen abundance here is believed to be the
first ever published for this galaxy.
The \ion{H}{I} extent is almost five times larger than the
optical extent \citep{cote00}.
The spectrum presented here was taken with the east-west 
slit placed so that the most northern star-forming clump
was observed.
The adopted oxygen abundance is 12$+$log(O/H) = 8.06
and the nitrogen-to-oxygen ratio is log(N/O) = $-1.91$.

\subsubsection{ESO302$-$G014 (AM0349$-$383)}

Little is described of this galaxy in the literature.
Because only \othreec\ and \halpha\ were detected in the spectrum,
there is no subsequent analysis.

\subsubsection{ESO347$-$G017, ESO348$-$G009}

Aside from the properties listed in Table~\ref{table_gxylist}, and
the inclusion of this galaxy in the \ion{H}{I} catalog by
\cite{cote97}, there is little in the literature describing further
these two dwarf galaxies in the Sculptor group. 
The adopted (bright-line) oxygen abundances, respectively, are:
12$+$log(O/H) = 7.84, and 7.98.
The nitrogen-to-oxygen ratios, respectively, are: 
log(N/O) = $-1.49$, and $-1.60$.
The neon-to-oxygen ratio for ESO347$-$G017 is log(Ne/O) = $-0.53$.

\subsubsection{ESO358$-$G060 (FCC 302)}
\label{sec_eso358g060}

This dwarf is a confirmed member of the Fornax Cluster 
\citep*[FCC 302;][]{drinkwater01,schroeder01}.
With an assumed distance modulus of 31.5 \citep{mould00}, the absolute
magnitude is $M_B \simeq -15.7$. 
The derived \ion{H}{I} mass to blue luminosity
ratio is $\simeq 4$, which is the largest in the present sample. 
However, the true distance is uncertain, which is a source of large
uncertainty in $M_{\rm HI}/L_B$. 

The derived oxygen abundance 
(12$+$log(O/H) = 7.32; $\sim$ 2.8\% of solar) makes this 
galaxy the most metal-poor dwarf in the present study.
The abundance is comparable to the second most metal-poor galaxy
known, i.e., the blue compact dwarf galaxy SBS~0335$-$052 with
12$+$log(O/H) = 7.33 \citep{melnick92,izotov97}.
If the abundance is truly low, additional spectroscopy of this dI with
a larger telescope should reveal \othreea, which should still be
observable even at the distance of $\sim 20$~Mpc; see
\cite{lee03virgo} for \othreea\ detections in dwarf galaxies at the
distance of the Virgo Cluster. 
The adopted nitrogen-to-oxygen ratio is log(N/O) = $-1.24$.

\subsubsection{IC 1613}

This is a well-studied dI in the Local Group whose position is near
the celestial equator. 
The $M_{\rm HI}/L_B$ value for IC~1613 is typical for its luminosity
compared to other dIs.
Its \ion{H}{II} regions have been catalogued by \cite{sandage71},
\cite{lequeux87}, \cite{price90}, and \cite{hlg90}.
The brightest \ion{H}{II} region was identified by Sandage as 
\#3 (S3), which has been labelled \ion{H}{II} \#37 by \cite{hlg90}.
\cite{talent80}, \cite{dr82}, and \cite{dk82} obtained photoelectric
spectrophotometry of S3, and all three studies showed that the
ionization source was a single Wolf-Rayet (W-R) star.
\cite{cole99} and \cite{dolphin01b} obtained resolved stellar
photometry and derived distances from the tip of the red giant branch
and the red clump, respectively; these distance measurements are in
agreement (730~kpc). 
\cite{rosado01} and \cite{vg01} obtained Fabry-Perot interferometry at
\halpha\ and \stwo\ to study the kinematics of S3 and S8 and found
that the nebular emission exhibits two-lobe structure with
superbubbles covering the entire galaxy.

\cite{talent80} and \cite{dk82} obtained \othreea\ measurements of
\ion{H}{II} \#37,
and derived 12$+$log(O/H) = 7.86 and 7.87, respectively, and 
log(N/O) = $-1.21$ (upper limit) and $-0.99$, respectively.
\cite{hg85} and \cite{pbtp88} measured additional spectrophotometry
for other \ion{H}{II} regions, of which one was identified
as a supernova remnant (\ion{H}{II} \#49, or Sandage \#8 or S8;
\citealp{dodb80,rosado01}).
For \ion{H}{II} \#49, \cite{dod83} derived 12$+$log(O/H) = 7.60 and
log(N/O) = $-0.90$, while \cite{pbtp88} derived 12$+$log(O/H) = 7.83
and log(N/O) = $-1.15$. 

Our spectrum of \ion{H}{II} \#37 (Fig.~\ref{fig_wr1613})
also exhibits broad Wolf-Rayet features near 4471,
4686, and 5800~\AA.  
Due to a strong \othreea\ detection, an oxygen abundance of
12$+$log(O/H) = $7.62 \pm 0.05$ was derived, which is the
adopted value here, and 1.6$\sigma$ lower than the \othreea\ abundance
(7.70) obtained most recently by \cite{kb95}.
Our log(N/O) = $-1.13 \pm 0.18$ value is in agreement with the 
value ($-0.96$) obtained by \cite{kb95}.
Our adopted neon-to-oxygen ratio is log(Ne/O) = $-0.60 \pm 0.05$,
which is in rough agreement with determinations by \cite{it99}
for other dwarf galaxies at similar oxygen abundance.
A new spectrum for \ion{H}{II} \#13 did not reveal \othreea.
Oxygen abundances and N/O values derived using the bright-line method
for \ion{H}{II} regions \#13 and \#37 are in agreement.


\subsubsection{IC 2032}

Not much is known about this dwarf, although it may be a member of the
Dorado group; see \cite{carrasco01} and references therein.
The \ion{H}{I} properties of the galaxy have been studied and
listed in the survey by \cite{hucht00a}.
There is a bright ``shell'' of star formation, giving the galaxy a
``cometary'' appearance.
The spectrum presented here was taken with the east-west slit placed
along the northern part of the shell, but \othreea\ was not detected.
The resulting (bright-line) oxygen abundance (12$+$log(O/H) = 7.97)
is the first ever published for this dwarf.
The adopted nitrogen- and neon-to-oxygen ratios are:
log(N/O) = $-1.37$, and log(Ne/O) = $-0.21$, respectively.
The derived Ne/O ratio is somewhat large compared to known values.
However, the $I(\nethree)/I(\hbeta)$ ratio is comparable to that
observed in the nearby dI IC~4662 in the Local Volume 
\citep{hmm90}.
Taking their published intensity ratios and temperatures for IC~4662,
the derived value of log(Ne/O) would be about zero.

\subsubsection{IC 5152}

%
Recent attention has been paid to this field dI, although a very
bright foreground star in the northwest corner of the galaxy prevents
deep imaging from taking place.
\cite{talent80} and \cite{webster83} obtained spectroscopy
of the brightest \ion{H}{II} region.
\cite{zm99} obtained deep stellar photometry and constructed
colour-magnitude diagrams; their field included the southeast
quadrant of the galaxy to avoid the bright foreground star.
A young population is indicated by the presence of \ion{H}{II} regions
and ultraviolet bright stars.
A distance was inferred after comparison with colour-magnitude
diagrams of other similar dIs and with theoretical isochrones.
For the intermediate-age stars a metallicity of one-tenth solar was
inferred.
With resolved photometry from the HST, \cite{kara02c} have recently
derived a TRGB distance of 2.07~Mpc.

The brightest \ion{H}{II} region is located at the northeast corner of
the dwarf and is labelled \ion{H}{II} region \#A by \cite{webster83};
this \ion{H}{II} region is believed to be the same one measured
by \cite{talent80}.
Our spectrum of \ion{H}{II} region \#A is the most up-to-date since
the work of \cite{talent80} and \cite{webster83}.  
\cite{talent80} measured \othreea\ and obtained an oxygen abundance 
of 12$+$log(O/H) = 8.36 and log(N/O) = $-1.52$.
\cite{webster83} derived a bright-line abundance of 
12$+$log(O/H) = 8.35.
We obtained an \othreea\ measurement of \ion{H}{II} region \#A and the
resulting oxygen abundance is 12$+$log(O/H) = $7.92 \pm 0.07$. 
Our value of the oxygen abundance is about one-tenth of the solar
value, which is consistent with the stellar metallicity inferred by
\cite{zm99}, but 2.7 times lower than the values reported
by \citeauthor{talent80} and \citeauthor{webster83}.
Our reported value of log(N/O) = $-1.05 \pm 0.12$ is three times
larger than the value given by \cite{talent80}.
Also, our adopted neon-to-oxygen ratio is log(Ne/O) = 
$-0.69 \pm 0.08$, which is in rough agreement with determinations by
\cite{it99} for other dwarf galaxies at similar oxygen abundance. 

Additional spectra have been obtained of an \ion{H}{II} region 
in the southwest region of the galaxy by \cite{hgo02}.
They measured \othreea\ for which they derived an oxygen abundance of
12$+$log(O/H) = 8.2, although their corrected $I(\halpha)/I(\hbeta)$
ratio is over a factor of two lower than the theoretical value for
typical \ion{H}{II} regions.
Future observations of this and other \ion{H}{II} regions would
be very helpful to confirm the nature of the unusual \ion{H}{II}
region observed by \citeauthor{hgo02}, the relatively large N/O, and
the homogeneity of oxygen abundances in this dwarf galaxy.
The ground-based stellar photometry obtained by \cite{zm99} includes
the centre and the southeast portion of the dwarf galaxy, which
unfortunately avoids the part of the galaxy where \ion{H}{II} region
\#A is located.
Deeper resolved photometry would be very useful in gaining further
clues about the underlying stellar populations and determining the
history of star formation.

\subsubsection{NGC 2915}

This galaxy is likely a relatively nearby blue compact dwarf
galaxy.
\cite{sersic77} presented one of the first comprehensive studies
using photoelectric photometry, spectrograms, and radio observations.
With updated broadband photometry, \cite{meurer94} found that the
dwarf contained two dominant stellar populations.
Current star formation is occurring at the centre where most
of the ionized gas and a bright blue population of stars are
present. 
A red diffuse population with an exponential surface brightness
profile lies outside of the central region.
\cite{meurer94} label this galaxy as an amorphous blue compact dwarf
galaxy with properties of a dwarf elliptical at large galactocentric
radii. 
\cite{meurer96} obtained \ion{H}{I} synthesis observations and found
that the \ion{H}{I} is five times larger in spatial extent than
the optical extent defined by the Holmberg radius. 
The galaxy also has a central bar and spiral arms with a maximum
rotation speed of $v_{\rm max} \simeq 85$ km~s$^{-1}$.
Their subsequent modelling shows that the dark matter content
is dominant at all radii.
\cite{bureau99} studied the barred spiral arm structure and
found that a rotating triaxial dark matter halo can best explain
the \ion{H}{I} observations.

The observed spectrum presented here was taken with the 
east-west slit placement through the centre of the galaxy. 
Our adopted (average) value of the oxygen abundance is
12$+$log(O/H) = 8.30.
The adopted (average) nitrogen-to-oxygen ratio is log(N/O) = $-1.48$.
From the spectrum of the galaxy nucleus measured by \cite{sersic77}, 
they found $I$(\halpha)/$I$(\othree) = 0.25 and 
[$I$(\halpha) + $I$(\ntwob)]/$I$(\stwo) = 5, which they claimed as
evidence for a low excitation \ion{H}{II} region.
Our corresponding ratios are: $I$(\halpha)/$I$(\othree) = 1.2, 
and [$I$(\halpha) + $I$(\ntwob)]/$I$(\stwo) = 5.1.
\cite{meurer94} also obtained a nuclear spectrum and derived an upper
limit to the oxygen abundance, 12$+$log(O/H) $\la$ 8.5, which is
consistent with our measurements.

\subsubsection{NGC 3109}
\label{sec_ngc3109}

NGC~3109 is the most massive galaxy in the nearby, extremely poor
Antlia-Sextans group \citep{vdb99,tully02}.
While considered as dI by some workers, this galaxy may be better
described as a Magellanic spiral or, perhaps, even a dwarf spiral
\citep{grebel01a,grebel01b}.
Studies of the \ion{H}{I} gas and stellar content have been well
documented for this galaxy.
For example, \cite{carignan85} and \cite{jc90} obtained optical and
\ion{H}{I} observations and showed that:
(1) compared to similar galaxies (e.g., SMC), the total galaxy
luminosity is comparable, but the optical extent is roughly two times
larger;
(2) the outer parts of the \ion{H}{I} gas is warped, likely due to 
an interaction with the Antlia dwarf galaxy;
(3) model fits to the rotation profile show that there is roughly
ten times more dark matter than luminous matter;
and (4) that the gas is a good tracer of the dark matter
distribution.
\cite{musella97} and \cite{minniti99} have obtained distances
of 1.36~Mpc and 1.33~Mpc using observations of Cepheid variable stars
and the tip of the red giant branch, respectively.
Also, \cite{kara02c} obtained a distance of 1.33~Mpc from the
magnitude of the tip of the red giant branch using HST WFPC2
observations.
\cite{grebel03} have obtained ${\rm [Fe/H]} \simeq -1.7$ dex for the
metallicity of the red giant branch, which points to the
metal-poor nature of this galaxy.

\cite{rm92} identified a number of \ion{H}{II} regions
and planetary nebulae in the eastern section of the galaxy.
M. McCall and C. Stevenson carried out spectroscopic measurements at
the Steward Observatory of \ion{H}{II} region ``\#5''; see
\cite{rm92} for their labelling.
An \othreea\ detection was reported and a lower branch 
oxygen abundance was obtained (7.73) by \cite{lee03}.
For the present work, measurements were carried out for \ion{H}{II}
region ``\#6'' (as labelled by \citealp{rm92}).
Using $I(\ntwob)/I(\halpha)$ and the calibrations by 
\cite{vanzee98sp} and \cite{denicolo02}, the resulting oxygen
abundances are consistent with lower branch values, but are
$\approx$ 0.4~dex higher than the values determined from either the
McGaugh or the Pilyugin calibration. 
We adopt here the McGaugh and Pilyugin calibrations, as all 
remaining bright-line oxygen abundances in the southern sample
are derived in a similar manner. 
So, taking the average of the ten values listed in
Table~\ref{table_nebabund}, the adopted oxygen abundance is 
12$+$log(O/H) = 7.63, which is similar to the value (7.73) determined
by \cite{lee03}.

\subsubsection{NGC 5264}
\label{sec_ngc5264}

NGC~5264 is a Cen~A dwarf galaxy with a TRGB distance of 4.53~Mpc
\citep{kara02b}.
Compared to other \ion{H}{II} spectra, our spectrum
shows relatively low $I(\othreec)/I(\hbeta)$,
whereas $I(\otwo)/I(\hbeta)$, $I(\ntwob)/I(\halpha)$, and
$I(\stwo)/I(\halpha)$ are high.
The \ntwootwo\ discriminant indicates an upper branch abundance.
Because of the low negative value ($-0.599$) for $\log\,O_{32}$,
the oxygen abundance derived using the Pilyugin bright-line method is
somewhat uncertain, as his method is calibrated to observations of
\ion{H}{II} regions with larger values of $\log\,O_{32}$.
Using $I(\ntwob)/I(\halpha)$, the calibrations of \cite{vanzee98sp}
and \cite{denicolo02} give 12$+$log(O/H) = 8.68 and 8.55,
respectively, which agree with the upper branch abundances from the
McGaugh and the Pilyugin calibrations.
Taking an average of the four values from each calibration, the
adopted oxygen abundance is 12$+$log(O/H) = 8.61, which is similar to
measured abundances in spiral galaxies.
Using the \cite{thurston96} method for \ion{H}{II} regions in spiral
galaxies, the nitrogen-to-oxygen abundance ratio is
log(N/O) = $-0.57$.

NGC~5264 has a higher oxygen abundance and a nitrogen-to-oxygen 
ratio than expected for its galaxy luminosity. 
This galaxy may be an example of a dwarf spiral (e.g.,
\citealp{schombert95}), whose luminosity is similar to the SMC, but
whose oxygen abundance is a factor of four larger than the SMC.  
A systematic spectroscopic survey of \ion{H}{II} regions in NGC~5264
should show whether or not the galaxy has a radial gradient in
oxygen abundance.

Compared to typical dIs, the $M_{\rm HI}/L_B$ value for NGC~5264 is
somewhat low compared to other late-type galaxies of similar
luminosity.
NGC~5264 is located one degree to the east of the luminous spiral
NGC~5236 (M~83), where an interaction with the latter could remove
\ion{H}{I} gas from NGC~5264. 
Interestingly, there is in the vicinity another Cen~A group member dI,
UGCA~365 (ESO444$-$G078), which appears to be even closer in
projection to M~83 (see Fig.~1 in \citealp{kara02b}).
There may be an interaction between M~83, UGCA~365, and
NGC~5264, which could explain the comparatively lower \ion{H}{I}
content in the latter galaxy.


\subsubsection{Sag DIG}

The Sagittarius dwarf irregular (Sag DIG) galaxy was first discovered
by \cite{cesarsky77} and later confirmed by \cite{longmore78}.
\cite{strobel91} detected what appeared to be three \ion{H}{II}
regions; only one appears to be a true \ion{H}{II} region, whereas the
other two sources are likely stars in \halpha\ emission.
\cite{yl97} found that the \ion{H}{I} component was more spatially
extended than the stellar component.
An \ion{H}{I} clump of relatively high density is nearly spatially
coincident with the \ion{H}{II} region.
There is no clear rotational motion for the \ion{H}{I} gas. 
Rather, the gas appears to be supported by random motions with
a broad- and a narrow-velocity component.
Using resolved stellar photometry, \cite{kara99} and \cite{lk00}
independently obtained TRGB distances of 1.11~Mpc, which places
this dwarf at the outer boundaries of the Local Group and makes
the membership to the Local Group uncertain.
Photometry has also revealed that the underlying stellar population is
very metal-poor with [Fe/H]~$\la -2$, which was also noted
by \cite{momany02}. 

\cite{stm89} attempted deep spectroscopy, but failed to detect
\othreea\ or \ntwob.
\cite{saviane02} have also recently obtained deep emission-line
spectroscopy of the \ion{H}{II} region.
Despite three hours of total exposure, \othreea\ was not detected.
Although \cite{saviane02} used a larger telescope, they did not detect
\othreea.
The oxygen abundance derived here (12$+$log(O/H) = 7.39)
agrees with the bright-line abundance obtained by \cite{stm89}
and with the upper end of the range of bright-line
abundances derived by \cite{saviane02}.
The oxygen-poor interstellar medium is consistent with the very low
iron abundance seen in the stars.
Although $\approx$ 0.2~dex lower than the value obtained by
\cite{saviane02}, our derived log(N/O) = $-1.63$ is almost identical
with the near-constant value for blue compact dwarf galaxies
\citep{it99}, even though Sag DIG is a very low-luminosity dI
with fairly quiescent star formation.
However, the $M_{\rm HI}/L_B$ value and the relatively low N/O
may be indicative of a very recent (and, perhaps, small) burst, which
is borne out by the young stars seen in resolved photometry
\citep{lk00}.

\subsubsection{UGCA 442}

\cite{miller96} obtained \halpha\ imaging of this edge-on 
galaxy in the Scl~group and found five \ion{H}{II} regions. 
Spectroscopy for two \ion{H}{II} regions in the southwest part
of the disk (\#2 and \#4 as labelled by \citealp{miller96}) did not
reveal \othreea; he derived an average bright-line oxygen abundance of
12$+$log(O/H) = 7.90.
These two \ion{H}{II} regions appear to be in a part of the
galaxy where \cite{cote00} found a local maximum in \ion{H}{I};
the \ion{H}{I} extent is about a factor of five larger than
the optical extent.
Our spectrum of \ion{H}{II} region \#2 (labelling by
\citealp{miller96}) also did not reveal \othreea,
and our subsequent bright-line oxygen abundance (12$+$log(O/H) = 7.85)
agrees with the value obtained by \cite{miller96}. 
The adopted nitrogen-to-oxygen ratio is log(N/O) = $-1.41$.
The $M_{\rm HI}/L_B$ value for this dwarf is comparable to typical
dIs at the given luminosity.

\subsection{Metallicity-Luminosity Relation}

\begin{figure*}
\centering
\includegraphics[width=9.cm]{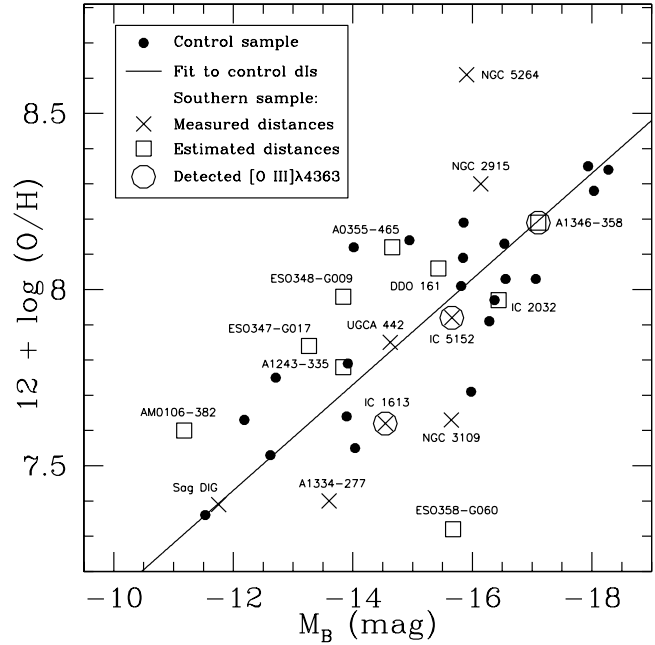} 
\caption{
Oxygen abundance versus galaxy luminosity in $B$.
Filled circles mark the control sample of dIs \citep{lee03}.
The best-fit to the control sample is shown as a solid line.
For the present sample of southern dwarfs, crosses and open squares
mark galaxies for which distances are measured and estimated,
respectively. 
In addition, galaxies for which \othreea\ has been measured are
marked with open circles.
For galaxies with \othreea\ measurements, the oxygen abundance
has an uncertainty of at most 0.1~dex.
The remaining galaxies without \othreea\ measurements have an
uncertainty of 0.2~dex in oxygen abundance.
A typical uncertainty of 0.2~mag in absolute magnitude accounts for
the various methods used to determine distances for dIs in both
samples.
}
\label{fig_oxylum}
\end{figure*}

\cite{skh89} and \cite{rm95} showed for dIs that metallicity
in the form of oxygen abundances increase with the galaxy luminosity 
in $B$.
This relationship has been interpreted as being representative of 
a relationship between metallicity and mass, at least where stellar
mass is concerned.
The metallicity-luminosity diagram is shown in Fig.~\ref{fig_oxylum}. 
Plotted are the set of dwarfs in the control sample from \cite{lee03}.
A new fit to galaxies with \othreea\ abundances and 
well-measured distances (i.e., control sample excluding NGC~3109, 
and new data for IC~1613 and IC~5152) leaves the slope and intercept
essentially unchanged within the stated errors compared
to the values obtained in the fits by \cite{rm95} and \cite{lee03}.

Plotted also are dwarfs from the southern sample, combining oxygen
abundances derived using the bright-line method and both 
measured and estimated distances.
Most dIs in the present sample have oxygen abundances consistent with
the metallicity-luminosity relation defined by dIs in the
control sample.
Because of the excellent agreement with the metallicity-luminosity
relation, IC~5152 will in the future be added to the growing list of
galaxies in the control sample with \othreea\ measurements and 
distances derived from resolved stellar photometry
(see Sect.~\ref{sec_control}). 

There may be an impression that the metallicity-luminosity relation
exhibits greater scatter at the range of luminosities shown.
At lower galaxy luminosities, Sag~DIG and A1334$-$277 have measured 
distances, but one should keep in mind that their oxygen abundances
were obtained using the bright-line method. 
ESO358$-$G060 is an outlier, which appears to be too bright in $B$
luminosity for its low oxygen abundance.
If the oxygen abundance is confirmed, this galaxy would resemble 
a blue compact dwarf galaxy, as galaxies representative of
this type lie mostly below the metallicity-luminosity relation 
\citep[e.g.,][]{ko00}.
At $M_B \approx -16$, the adopted oxygen abundance for NGC~5264
(Sect.~\ref{sec_ngc5264}) is approximately 0.2 to 0.3~dex higher than
expected at the given luminosity.
Additional \ion{H}{II} region spectra would be useful in
confirming this result.


\subsection{Relative Nitrogen to Oxygen Abundances}

\begin{figure*}
\centering
\includegraphics[width=9.cm]{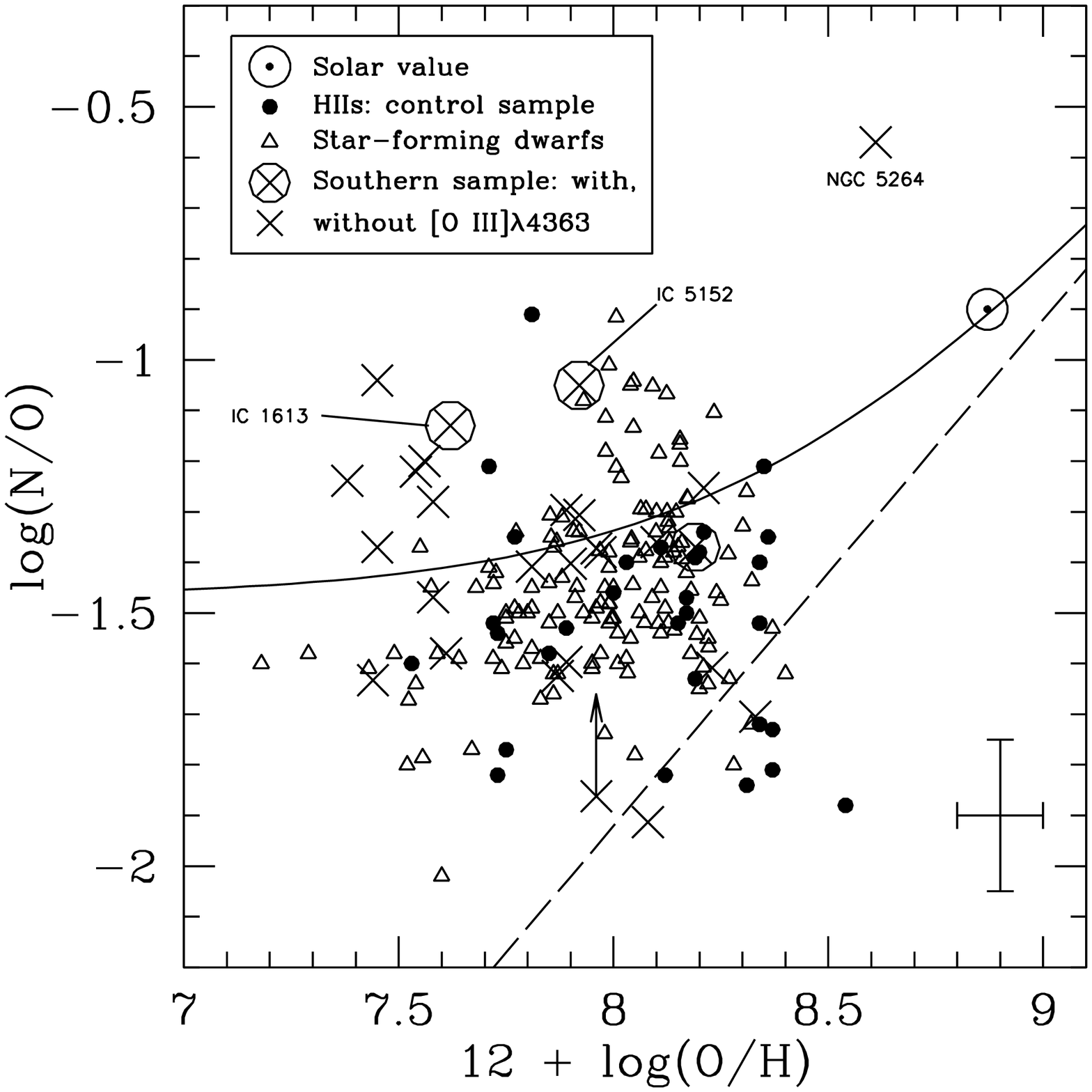} 
\caption{
Nitrogen-to-oxygen abundance ratio versus oxygen abundance.
Filled circles indicate \ion{H}{II} regions from the control sample of
dIs, whose oxygen abundances were obtained directly from measurements
of the \othreea\ emission line.
Open triangles represent other star-forming dwarf galaxies from
\cite{garnett90}, \cite{ks96}, \cite{vanzee97}, and \cite{it99}.
Crosses mark \ion{H}{II} regions from dwarfs in the southern sample;
open circles denote galaxies where \othreea\ was detected.
Labelled are IC~1613 \ion{H}{II}\#37 and IC~5152 \ion{H}{II}\#A
for their unusually high N/O ratios.
Two models for the production of nitrogen are shown \citep{vce93};
the solid line is a model with both primary and secondary production,
whereas the dashed line is a model for secondary production.
The solar value of N/O \citep{zsolar} is also indicated.
Typical errors in log(O/H) and log(N/O) are shown at the
lower right. 
}
\label{fig_nooxy}
\end{figure*}

A plot of log(N/O) versus log(O/H) is shown for star-forming dwarf
galaxies in Fig.~\ref{fig_nooxy}.
In general, the present data overlaps with the loci defined
by dwarfs in the control sample \citep{lee03} and by other dwarf
galaxies \citep{garnett90,ks96,vanzee97,it99}, although there are a
few \ion{H}{II} regions with elevated values of log(N/O) for their 
oxygen abundance. 
There still appears to be a great deal of dispersion in N/O
values at a given oxygen abundance.
For IC~1613 \ion{H}{II}\#37 and IC~5152 \ion{H}{II}\#A,
their log(N/O) values are somewhat elevated compared to 
the mean value for \ion{H}{II} from dIs in the control sample.
For NGC~5264, the oxygen and nitrogen abundances are in better
agreement with values found in spiral galaxies;
see the compilation by \cite{henry00}. 

\section{Conclusions}		
\label{sec_concl}		

Optical emission-line spectroscopy of \ion{H}{II} regions
was obtained in dwarf irregular galaxies in the
Centaurus~A group, the Sculptor group, the Antlia-Sextans group,
the Local Group, and in the field.
\othreea\ was measured in A1346$-$358, IC~1613, and IC~5152, and
the standard method was used to obtain oxygen abundances.
For the remaining galaxies, the bright-line method was used to compute
oxygen abundances with the McGaugh and Pilyugin calibrations.
The \ntwootwo\ intensity ratio was used to break the degeneracy
in the bright-line method and applied to both calibrations.
For NGC~3109 and NGC~5264, the [N~II]/\halpha\ intensity ratio
was also used to confirm their abundances.
ESO358$-$G030 has the lowest oxygen abundance in the sample
with a value of 12$+$log(O/H) = 7.32, which is comparable
to the value for the second most metal-poor galaxy known
(SBS~0335$-$052).

Oxygen abundances for dwarfs in the southern sample agree with the 
metallicity-luminosity relationship defined by dwarf irregulars
in the control sample.
Three dwarfs with direct abundances are found to be in very good
agreement with the relation, whereas the remaining galaxies with
bright-line abundances are in comparable agreement with 
larger scatter.
NGC~5264 appears to have an oxygen abundance approximately 
two times larger than expected for its luminosity.
Extensive spectroscopy of \ion{H}{II} regions in NGC~5264 will confirm
whether the galaxy is a dwarf irregular galaxy with a zero or very
small abundance gradient or a low-luminosity spiral galaxy with a
significant abundance gradient.

Nitrogen-to-oxygen abundance ratios for dwarfs in the southern sample
are comparable to those in the control sample at a given oxygen
abundance, although IC~1613 and IC~5152 have larger N/O values
compared to nearby dwarf irregulars at a given oxygen abundance.
Resolved stellar photometry shows that IC~1613 is currently undergoing
a lower rate of star formation activity than in the past, but the
dwarf has had a constant, low level of star formation activity
throughout its history. 
Additional stellar photometry for IC~5152 should provide further clues
to the star formation and enrichment history of this nearby dwarf.
%

\begin{acknowledgements}	

We are grateful to the anonymous referee for improving the
presentation of the manuscript, and to the CTIO staff for their help
with the observations.
Partial support for this work was provided by NASA through grant 
GO-08192.97A from the Space Telescope Science Institute, which is 
operated by the Association of Universities for Research in 
Astronomy, Inc., under NASA contract NAS5-26555.
H. L. acknowledges MPIA for their financial support, 
and Stephanie C\^ot\'e and Evan Skillman for discussions and 
making available a copy of their paper before publication.
Some data were accessed as Guest User, Canadian Astronomy Data Center,
which is operated by the Dominion Astrophysical Observatory for the 
National Research Council of Canada's Herzberg Institute of Astrophysics. 
This research has made use of the NASA/IPAC Extragalactic Database
(NED), which is operated by the Jet Propulsion Laboratory,
California Institute of Technology, under contract with the
National Aeronautics and Space Administration. 
\end{acknowledgements}

\appendix

\section{Other Measurements}
\label{sec_app}

\cite{scm03} have obtained emission-line spectroscopy of late-type
dwarf galaxies in the Sculptor group with the 4-m telescope at CTIO.
Their sample included ESO347$-$G017, ESO348$-$G009, and UGCA~442;
they measured \othreea\ in ESO347$-$G017 and UGCA~442. 
Our bright-line oxygen abundances for ESO347$-$G017,
ESO348$-$G009, and UGCA~442 agree with the abundances
obtained by \citeauthor{scm03}


%
\end{document}